\documentclass[12pt]{iopart}

\usepackage{iopams,amssymb}
\usepackage{amssymb,amsfonts}
\usepackage{graphicx}
\graphicspath{{Figures/}}
\usepackage{psfrag}
\usepackage[usenames,dvipsnames]{xcolor}
\def\contentsname{}
\usepackage{bm}

\newcommand{\trho}{\tilde\rho}
\newcommand{\txi}{\tilde\xi}

\begin{document}
\bibliographystyle{unsrt}

\title{Multi-Scale  Exciton and Electron Transfer in Multi-Level Donor-Acceptor System}

\author{Shmuel Gurvitz}
 \ead{shmuel.gurvitz@weizmann.ac.il}

\address{Department of Particle Physics and Astrophysics, Weizmann
Institute, 76100, Rehovot, Israel }
\ead{shmuel.gurvitz@weizmann.ac.il}

\author{Alexander I. Nesterov }

\address{Departamento de F{\'\i}sica, CUCEI, Universidad de Guadalajara,
Av. Revoluci\'on 1500, Guadalajara, CP 44420, Jalisco, M\'exico}
\ead{nesterov@cencar.udg.mx}

\author{Gennady P.  Berman}

\address{Theoretical Division, T-4, Los Alamos National Laboratory, and the New Mexican Consortium, Los Alamos, NM 87544, USA}
 \ead{gpb@lanl.gov}

\begin{abstract}
We  study theoretically the noise-assisted quantum exciton  (electron) transfer (ET) in bio-complexes consisting of a single-level electron donor and an acceptor which has a complicated internal structure, and is modeled by many electron energy levels. Interactions are included between the donor and the acceptor energy levels and  with  the protein-solvent noisy environment. Different regions of parameters are considered, which characterize (i) the number of the acceptor levels, (ii) the acceptor ``band-width", and (iii) the amplitude of noise and its correlation time. Under some conditions, we derive  analytical expressions for the ET rate and efficiency. We obtain equal occupation of all levels at large times, independently of the structure of the acceptor band and the noise parameters, but under the condition of non-degeneracy of the acceptor energy levels. We discuss the multi-scale dynamics of the acceptor population, and the accompanying effect of quantum coherent oscillations. We also demonstrate that for large number of levels in the acceptor band, the efficiency of ET can be close to 100\%, for both downhill and uphill transitions and for sharp and flat redox potentials.
\end{abstract}

\pacs{ 87.15.ht, 05.60.Gg, 82.39.Jn}


\maketitle

\section{Introduction}

Exciton (and electron) transfer (ET) in photosynthetic bio-complexes, including plants, eukaryotic algae and cyanobacteria, has a large range of characteristic time-scales, from tens of femtoseconds to milliseconds and more.
In particular, ET in light-harvesting complexes (LHCs) and primary charge separation processes in the reaction centers (RCs) of the photosystem I (PSI) and  photosystem II (PSII) take place on very short time-scales, of order $1-5 \rm ps$.
    It was recently experimentally discovered  that the exciton dynamics in this type of systems can involve quantum coherent effects (quantum Brownian motion) \cite{book,Len,ECR,CWWC,PHFC}.
These results have generated significant interest in creating adequate mathematical tools for describing and modeling of quantum coherent processes in these systems \cite{IF,IFG,IF2,RMKL,CFMB,MBS}.

When considering analytically the ET dynamics  in LHCs, usually one uses the F\"{o}rster resonant perturbation theory with different generalizations \cite{May,Forster1,Forster2}.  In spite of this approach, based on application of the Fermi's Golden Rule (FGR),  is very straightforward, physically visible, and  useful in many applications, it does not  describe the multi-scale ET dynamics, which usually is the case for donor and acceptor with  the {\it finite} energy  band-widths. (See our results below.) Indeed,  a single ET rate, $\Gamma_{FGR}=2\pi |V|^2\rho/\hbar$, which does not depend on time and formally is valid for $t\in [0,\infty]$,  occurs  in the well-known Weisskopf-Wigner model \cite{SM},  where $V$ is the matrix element of interaction between a single  electron energy  level (donor)  with the acceptor, modeled by an infinite  energy band with the density of homogeneously distributed electron states,  $\rho$. In the case of the initially populated donor, the acceptor is  populates in time  with the probability: $P_A(t)=1-\exp(-\Gamma_{FDR}t)$. Generally, the FGR approach and, based on it the F\"{o}rster resonant perturbation theory, are valid only for some region of parameters and for some intermediate times, and don't describe the multi-scale ET dynamics.

Then, it becomes an important issue  to develop an approach which (i) can describe analytically  the ET rates in the LHCs on different time-scales,  (ii)  is valid for finite energy bands of both donor and acceptor, and (iii) can easily be  implemented in numerical analysis. The approach, developed below, contributes in resolution of these issues, which is the main subject of our paper.

In real situations, the LHCs and RCs networks consist of many chlorophyll, carotenoids, and other complex organic molecules which include the corresponding electron and exciton quantum states with different energy levels. So, for modeling of the electron and exciton transfer in these complexes, an extremely large Hilbert space should be used. Then, to make the corresponding models useful and predictive, different parts of the LHCs and RCs  can be considered, under some conditions, as interacting clusters with particular sets of molecules, geometries, and structures of exciton and electron energy levels. This approach, to some extent, is equivalent to the well-known coarse-graining procedure, which is usually used for these purposes. (See, for example \cite{BAF}, and references therein.) Then, the questions arise: How do the structures of individual clusters, the interactions between different clusters, and their interactions with noisy or thermal protein environment(s) affect the electron and exciton transfer?

In this paper, the ET in a simple donor-acceptor (two-cluster) model is considered, which allowed us to derive many useful preliminary results in these directions. Namely, we analytically and numerically study the ET  between a single-level electron donor interacting with an acceptor, which is modeled by a finite number of electron energy levels.
In particular, this situation takes place when a single excited energy level (say, in $Chlb$) is initially populated, and  the acceptor (excited electron state of $Chla$) has a complex structure due to the contribution of the $Chla$ vibrational levels \cite{Fulton,Lutz,But,Scholes,F1,F2,F3}.

Similar situation takes place, for example, when a coarse-graining procedure can be applied, and a bio-complex with many energy levels can be considered as an electron ``donor" and/or an ``acceptor" with complicated internal structures.   We assume that an external (classical) diagonal noise  interacts with both the donor and  the acceptor energy levels. This approach is often used for modeling the protein-solvent environment under non-equilibrium conditions. (See, for example, \cite{Dewey,Ben,Skourtis,Das,Nesterov1,NBSS}, and references therein.)
Note, that our model can be easily generalized to include (i) both multi-level donor and acceptor,  (ii) complex band structures, (iii) interactions inside the bands, (iv) more than two interacting donor-acceptor type clusters, (v) thermal (instead of noisy) environment(s).

Our main results include:

1) We derived analytical expressions for the ET rates and efficiencies in different regions of parameters, which are  important for understanding a complicated quantum dynamics of the system. We also provided the numerical simulations which confirm our analytical results.

2) We showed that generally the dynamics of the acceptor population is characterized by multi-scale processes, accompanied by quantum coherent oscillations. We estimated the corresponding ET rates and the period of these oscillations.

3) We obtained the equal occupation of all levels at large times, independent of the structure of the acceptor band, but under the condition of non-degeneracy of the acceptor energy levels. The case of a degenerate acceptor band is analyzed in details.

4)	We demonstrated that the efficiency of population of the acceptor can be close to 100 \%, in relatively short times, for both sharp and flat redox potentials.

5) We demonstrated the possibility of the efficient uphill population of the acceptor, due to the ``entropy factor" (large number of levels in the acceptor band).

The paper is organized as follows. In Sec. II, we consider a single-level donor interacting with a multi-level acceptor, in the absence of noise. Some general features of this system are described for this case. In Sec. III, we introduce an approach based on the master equation for the reduced density matrix, in the presence of noise. In Sec. IV, we present  analytical and numerical results for the ET rates for different regions of parameters, for a single-level donor and single-level acceptor system. In Sec. V, we extend the results to the $N$-level acceptor. In Appendix A, we derive the mathematical expressions needed in Sec. V.  In the Conclusion, we summarize our results, and discuss possible generalizations of our approach.

\section{Single-level donor interacting with multi-level acceptor}

\begin{figure}[tbh]
\begin{center}
\scalebox{0.75}{\includegraphics{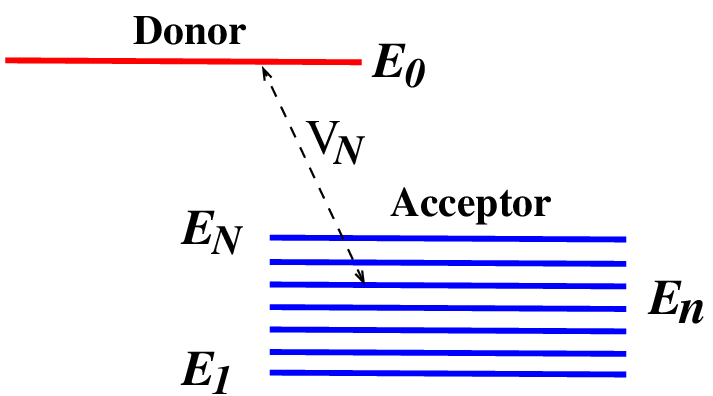}}
\end{center}
\caption{(Color online) Schematic of our simplified model consisting of an single-level donor and of a $N$-level  acceptor.
\label{S1a}}
\end{figure}

Consider a single level donor coupled with a $N$-level band of acceptor. This is shown schematically in Fig. \ref{S1a}.
We describe this system by the following tunneling Hamiltonian,
\begin{eqnarray} \label{H1}
H_S = E_0|0\rangle\langle 0|  +  \sum_{n=1}^N E_{n}|n\rangle\langle  n |+\sum_{n=1}^NV_N\big ( |n\rangle\langle  0| + |0\rangle\langle n |\big),
\end{eqnarray}
where $E_0$ is the energy level of the donor and $E_n$ is the $n$-th level of the acceptor-band. We assume that the interaction (tunneling coupling), $V_N$, is the same for all levels of the acceptor, but depends on the number of levels ($N$) as
$V_N=V/\sqrt{N}$. This dependence results from the normalization factor of the acceptor states, $|n\rangle$. Note, that the same Hamiltonian describes either the exciton (energy) transfer or the electron tunneling \cite{silbey}, and therefore the corresponding dynamics will be the same.

Consider the donor, $|0\rangle$, is initially populated by an exciton. Due to its coupling with the acceptor, the exciton makes transitions to the acceptor's states, described by the time-dependent wave function, $|\Psi(t)\rangle$. The latter can be written as,
\begin{eqnarray}
|\Psi(t)\rangle =b_0(t)|0\rangle +\sum_{n=1}^Nb_n(t)|n\rangle,
\label{wf}
\end{eqnarray}
with the initial condition, $|\Psi (0)\rangle =|0\rangle$.
Substituting Eq.~(\ref{wf}) into the Schr\"odinger equation,
$i\partial_t|\Psi(t)\rangle =H_S|\Psi(t)\rangle$, we obtain \cite{mil},
\begin{eqnarray}
&i\dot b_0(t)=E_0b_0(t)+\sum_{n=1}^N V_Nb_n(t)\label{a1a},\\
&i\dot b_n(t)=E_nb_n(t)+V_Nb_0(t).
\label{a1b}
\end{eqnarray}
(Here and below we choose $\hbar=1$.) In order to solve these equations, we apply the Laplace transform, $\tilde b(E)=\int_0^\infty b(t)e^{iEt}dt$.
As a result we obtain,
\begin{eqnarray}
&(E-E_0)\tilde b_0(E)-\sum_{n=1}^N V_N\tilde b_n(E)=i\label{a3a},\\
&(E-E_n)\tilde b_n(E)-V_N\tilde b_0(E)=0.
\label{a3b}
\end{eqnarray}
The r.h.s. of these equations reflects the initial condition.

Substituting $\tilde b_n(E)$ from the second equation into the first one, we obtain for the amplitude, $b_0(E)$,
\begin{eqnarray}
\tilde b_0(E)={\displaystyle i\over\displaystyle E-E_0-\sum_{n=1}^N {V_N^2\over E-E_n}}.
\label{a2}
\end{eqnarray}
The time-dependent amplitude of the donor population, $b_0(t)$, is given by the inverse Laplace transform,
\begin{eqnarray}
b_0(t)={1\over 2\pi}\int\limits_{-\infty}^\infty \tilde b_0(E)\,e^{-iEt}\,dE.
\label{invlap}
\end{eqnarray}
The probability of finding the donor occupied (survival probability) is, $P_0(t)=|b_0(t)|^2$. Respectively, the occupation probability of the acceptor' $n$-th level, $P_n(t)=|b_n(t)|^2$, is obtained through the inverse Laplace transform of the amplitude, $\tilde b_n(E)=V_N\,b_0(E)/(E-E_n)$, Eq.~(\ref{a3b}).

For the infinite acceptor' band ($E_1\to -\infty$,  $E_N\to \infty$, and $N\to\infty$) we can replace, $\sum_n\to\int\varrho_N^{}(E_n)dE_n$, where $\varrho_N^{}(E_n)$ is the density of states. It can be written as, $\varrho_N^{}(E_n)=\varrho\, N$, where we assume that $\varrho$ is independent of $E_n$.
Then we obtain,
\begin{eqnarray}
\sum_n {V_N^2\over E-E_n}\to \int\limits_{-\infty}^{\infty}
{V_N^2\varrho\, N\,dE_n\over E_n-E}=i\pi V^2\varrho.
\end{eqnarray}
Substituting this result into Eqs.~(\ref{a2}) and (\ref{invlap}), one can easily perform integration over $E$ by closing the integration contour in the lower half-plane. As a result, $|b_0(t)|^2=\exp (-\Gamma t)$, where $\Gamma =2\pi V^2\varrho$. Therefore, the donor is totally depopulated in the limit of $t\to\infty$,  in agreement with the Weisskopf-Wigner approach \cite{SM}.

This is not the case for a finite $N$. In particular, when the donor' energy is far outside the acceptor band, $|E_0-E_N|\gg V$, the transition probability is very small, so the donor remains almost totally populated in the limit of $t\to\infty$. This can be easily illustrated by the example of $N=1$. Evaluating the integral (\ref{invlap}) for this case, we find for occupation of the donor,
\begin{eqnarray}
P_0(t)={\epsilon^2+4V^2\cos^2\omega_R^{}t\over \epsilon^2+4V^2}
\label{rabi}
\end{eqnarray}
where $\epsilon =E_0-E_1$, and $\omega_R^{}={1\over2}\sqrt{4V^2+\epsilon^2}$ is the frequency of Rabi oscillations. It follows from Eq.~(\ref{rabi}) that the occupation of donor at $t\to\infty$ oscillates in time and is proportional to $4V^2/\epsilon^2\ll 1$.

The same occurs for any finite $N$, and when $|E_0-E_N|\gg V$. Indeed, solving Eqs.~(\ref{a3a})-(\ref{a3b}) by iterations (in powers of interaction), we find,
\begin{eqnarray}
&\tilde b_0(E)={i\over E-E_0}+\cdots\nonumber\\
&\tilde b_n(E)={iV_N\over (E-E_0)(E-E_n)}+\cdots.
\label{pert}
\end{eqnarray}
Then, performing the inverse Laplace transform, Eq.~(\ref{invlap}), we obtain for the acceptor' population, $P_a(t)$,
\begin{eqnarray}
P_a(t)=\sum_{n=1}^N|b_n(t)|^2\sim N{V_N^2\over\epsilon^2}\sim {V^2\over \epsilon^2}\ll 1.
\label{pert1}
\end{eqnarray}

\begin{figure}[tbh]
\begin{center}
\scalebox{0.75}{\includegraphics{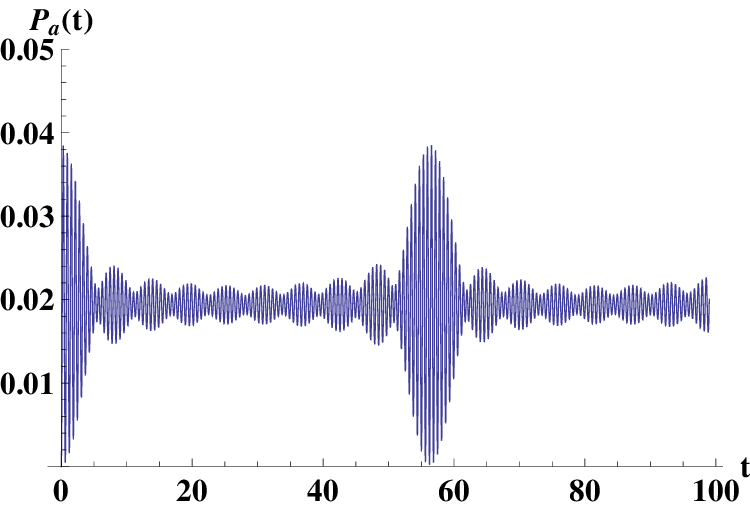}}
\end{center}
\caption{(Color online) Occupation of the acceptor as a function of time; $N=10$, $\epsilon =10$, $V=1$, and the acceptor's band-width, $\delta_a=1$.
\label{fig2}}
\end{figure}

An example of the acceptor' population, obtained from the exact numerical solution of Eqs.~(\ref{a1a})-(\ref{a1b}), is shown in Fig.~\ref{fig2} for $N=10$ and for acceptor' levels being equally distributed inside the band,
\begin{eqnarray}
E_n={2n-1-N\over 2(N-1)}\,\delta_a,
\label{acen}
\end{eqnarray}
where $\delta_a$ is the band-width of the acceptor. The energy of the band-center is set to zero, so that $E_1=-\delta_a/2$,  $E_N=\delta_a/2$ and  the donor's energy is $E_0=\epsilon$.

The parameters used in Fig.~\ref{fig2} correspond to $\delta_a=1$, $V=1$, and $\epsilon =10$, in arbitrary units. (For instance, it is convenient to choose the energy parameters in units of $ps^{-1}$, where time is measured in $ps$. Then, the values of our parameters in energy units should be multiplied values by $\hbar\approx 6.58\times 10^{-13}\rm meVs$, so that $\epsilon = 10\; \rm ps^{-1}\approx 6.58\rm meV$).

One finds that the results shown in Fig.~\ref{fig2}
confirm our estimates,  Eq.~(\ref{pert1}),  based on the perturbative calculations (assuming small $V$), that the acceptor cannot be populated if the energy levels of the donor are outside the acceptor band. This situation can change very drastically in the presence of noise.

\section{Density matrix approach and inclusion of noise}

Let us consider the donor and acceptor interacting with an external environment that perturbs their energy levels. The total Hamiltonian can be written as,
\begin{eqnarray}
H=H_S+H_{env}+H_{int},
\label{env}
\end{eqnarray}
where $H_S$ is the Hamiltonian of the donor-acceptor system, Eq.~(\ref{H1}), and the two other terms describe the environment and its interaction with the system, correspondingly. By solving the Schr\"odinger equation, $i\partial_t|\Psi (t)\rangle =H|\Psi (t)\rangle $, we can find the total wave function, $|\Psi(t)\rangle$. In order to determine how the environment affects the system, we take the trace of the total density matrix  over all variables of the environment, $\rho(t) ={\rm Tr}_{env} \left[|\Psi (t)\rangle\langle\Psi(t)|\right]$. The resulting (reduced) density matrix describes the system' behavior under the influence of the environment.

Now we specify the interaction term. We assume it as a product of two operators, $H_{int}={\cal V}{\cal U}_{env}$, where ${\cal  V}$ acts on the system and ${\cal U}_{env}$ on the environment. One can find many examples of this interaction. For instance, the electron of the system can be coupled capacitively to a nearby fluctuator, generated by a current flowing through it (as in a single-electron transistor), or by any other mechanism \cite{GABS,GM}, belonging to the environment. In biological systems, when considering exciton and electron transfer, noise is mainly caused by the protein environment \cite{Dewey,Ben,Skourtis,Das,Nesterov1,NBSS}. Then, one can write for the ET,
\begin{eqnarray}
{\cal V}{\cal U}_{env}=\left(U_d|0\rangle\langle 0|+U_a\sum_{n=1}^N|n\rangle\langle n|\right)c^\dagger_{}c,
\label{noise1}
\end{eqnarray}
where $U_{d,a}$ is the Coulomb interaction between the electron on the donor (or on the acceptor) and the fluctuator, and $c^\dagger_{}c$ is the electron density operator of the fluctuator. Note, that when applying our approach to a donor-acceptor system which describes the exciton transfer in bio-complexes, instead of the Coulomb interaction, the dipole-dipole interaction is usually used. In this case, instead of the density operator, $c^\dagger_{}c$, in Eq.~(\ref{noise1}), a bosonic operator of the type, $\varphi=\sum_k(g_ka_k^\dagger+h.c.)$, is used (see, for example \cite{MBS,HDR}, and references therein), where $g_k$ is a form factor related to the protein environment, and $a_k^\dagger$ and $a_k$ are the creation and annihilation operators of the $k$-th mode.

We assume that $U_d\not =U_a$, since the donor and the acceptor are at different spatial locations. This model describes telegraph noise, because $c_{}^\dagger c$ can have only two values: $1$ or $0$ for an occupied or  unoccupied fluctuator. Since the effect of noise on the system is not sensitive to the particular microscopic origin of noise, we restrict our study to this model only. This allows us to investigate the effect of the environment on the system in the simplest way. The problems of telegraph noise have been considered in many publications \cite{GABS,GM,Blume,Tokura,BGA,Gurvitz1}. We choose the approach of Ref. \cite{Gurvitz1}, which results in Bloch-type master equations, which have a very transparent physical meaning. In addition, these equations can be rigorously derived for certain microscopic models of telegraph noise \cite{GM}, or by averaging the time-dependent fluctuating energy levels, by using a very effective method of Shapiro and Loginov \cite{shap,gae}.

In this paper, we do not present a detailed microscopic treatment of these equations for a particular $H_{env}$, but simply replace the operator $c^\dagger_{}c$ by a random variable, $[\zeta(t) +1]/2$, where $\zeta(t)$ jumps randomly from 1 to -1 (or from $-1$ to $1$)  at a rate $\gamma_+$ (or $\gamma_-$)  \cite{Blume}. This procedure is valid whenever there is no back-action from the system on the noise spectrum \cite{GM}.

The noise distribution is determined by the probabilities, $p_{\pm}^{}$, of finding $\zeta(t)$ at the values $\pm 1$. Detailed balance then implies the relation, $p_-^{}\gamma_+^{}=p_+^{}\gamma_-^{}$, and therefore $p_\pm^{}=\gamma_{\pm}^{}/(2\gamma)$, where $\gamma=(\gamma_+^{}+\gamma_-^{})/2$ is the inverse time associated with noise ($\gamma =2/S(0)$, where $S(\omega)$ is the noise spectrum \cite{Gurvitz1}).

The symmetric noise, $\gamma_+=\gamma_-$, corresponds to $p_+=p_-$, and therefore it formally corresponds to the infinite temperature regime. Respectively, an asymmetric noise, $\gamma_+\not =\gamma_-$, would be considered as one corresponding to the finite temperature regime (see, for example, \cite{HDR,MRSN,Capek}.) However, we have to point out that in our consideration noise is not an equilibrium sub-system. Indeed, noise we are dealing with is sustained in its steady state by an external source and its spectrum, and it is not affected by the  system. The latter, therefore cannot be in a thermal equilibrium with the noise, as is demonstrated explicitly below.

It is interesting to compare our treatment of noise with that of Haken, Reineker and Strobl \cite{Capek,GHSR1,GHSR2,GHSR3}, which on the first sight look similar. However, the main assumption of their treatment is that the  phonon dynamics is infinitely quick as compared to that of the exciton. This implies that noise is $\delta$-correlated in time \cite{Capek}. This is not the case of our approach, which has not such restrictions.  Moreover, most important effects of noise in exciton (electron) transport take place when noise dynamics is comparable with that of the exciton (electron) one \cite{gae}. Another, interesting descriptions of random effects in the ET have been proposed by use of the Random Matrix Theory technique and Keldysh non-equilibrium Green's function approach \cite{Gud,Bih}.

Although we use a particular model for the noise, the main goal of our approach is to demonstrate analytically and numerically the multi-scale ET dynamics, and not the dependencies on the characteristic parameters of an external reservoir. Indeed, in  this context, the main function of the external noise in our approach is to assist the ET when the donor level (band) does not overlap with the acceptor band.  We also note that because the ET rates in LHCs are relatively large ($\sim$ ps$^{-1}$, or even more), no consensus exists on what are the main contributions in the non-equilibrium ET dynamics (with non-zero reduced density matrix elements), noise or thermal fluctuations, or both.

Let us obtain these equations for the reduced density matrix of the system, $\rho (t)$. First consider the case of no interaction with the environment, $H_{int}=0$. Then,  the density matrix of the system, defined as
$\rho_{jj'}(t)=b_j^{}(t)b_{j'}^*(t)$, satisfies the following Bloch-type equations,
\begin{eqnarray}
&\dot\rho_{00}=i V_N\sum_{n=1}^N\big(\rho_{0n}-\rho_{n0}\big),
\label{b1a}\\
&\dot\rho_{nn'}=i(E_{n'}-E_n)\rho_{nn'},
+i V_N\big(\rho_{n0}-\rho_{0n'}\big),
\label{b1b}\\
&\dot\rho_{0n}=i(E_n-E_0)\rho_{0n}
+i V_N\big(\rho_{00}-\sum_{n'}\rho_{n'n}\big).
\label{b1c}
\end{eqnarray}
These equations are derived straightforwardly from Eqs.~(\ref{a1a})-(\ref{a1b}).

In the presence of interaction, Eq.~(\ref{noise1}), the energy levels of the donor and acceptor in Eqs.~(\ref{b1a})-(\ref{b1c}) are replaced by $E_{0,n}\to E_{0,n}+{1\over2}U_{d,a}\zeta(t)$, where
the constant terms, $U_{d,a}/2$, were included in the definition of the energy levels. As mentioned above, in the case of the exciton energy transfer, the operator, $c^\dagger_{}c$, should be replaced by the bosonic operator, $\varphi$, which in our case is reduced to a random variable, $\zeta(t)/2$, with the ensemble average, $\overline\zeta(t)=0$. So, the constant terms, $U_{d,a}/2$, do not appear in the renormalization of the donor and acceptor energy levels.

Now it is quite natural to extend Eqs.~(\ref{b1a})-(\ref{b1c}) to include noise by replacing the density matrix $\rho_{jj'}^{}(t)$, Eqs.~(\ref{b1a})-(\ref{b1c}), by the two-component vector,  $\{\rho_{jj'}^{(+)}(t),\rho_{jj'}^{(-)}(t)\}$. For the redox potential (energy gap),  $E_0-E_n\to E_0-E_n \pm D$ in Eq.~(\ref{b1c}), where $D=(U_d-U_a)/2$. In addition, the stochastic hopping terms, $\rho_{jj'}^{(+)}(t)\longleftrightarrow\rho_{jj'}^{(-)}(t)$, with rates, $\gamma_{\pm}^{}$, should be included in the equation of motion. Thus, we replace Eqs.~(\ref{b1a})-(\ref{b1c}) by the following equations of motion, which now include noise,
\begin{eqnarray}
&\dot\rho_{00}^{(+)}=iV_N\,\sum_{n=1}^N\big(\rho_{0n}^{(+)}
-\rho_{n0}^{(+)}\big)
-\gamma_-^{}\rho_{00}^{(+)}+\gamma_+^{}\rho_{00}^{(-)},
\label{a11a}\\
&\dot\rho_{00}^{(-)}=iV_N\,\sum_{n=1}^N\big(\rho_{0n}^{(-)}
-\rho_{n0}^{(-)}\big)
-\gamma_+^{}\rho_{00}^{(-)}+\gamma_-^{}\rho_{00}^{(+)},
\label{a11b}\\
&\dot\rho_{nn'}^{(+)}=i(E_{n'}-E_n)\rho_{nn'}^{(+)}
+iV_N\,\big(\rho_{n0}^{(+)}
-\rho_{0n'}^{(+)}\big) \nonumber \\
&-\gamma_-^{}\rho_{nn'}^{(+)}+\gamma_+^{}\rho_{nn'}^{(-)},
\label{a11c}
\end{eqnarray}
\begin{eqnarray}
&\dot\rho_{nn'}^{(-)}=i(E_{n'}-E_n)\rho_{nn'}^{(-)}
+iV_N\,\big(\rho_{n0}^{(-)}
-\rho_{0n'}^{(-)}\big)
-\gamma_+^{}\rho_{nn'}^{(-)}+\gamma_-^{}\rho_{nn'}^{(+)},
\label{a11d}\\
&\dot\rho_{0n}^{(+)}=i(E_n-E_0-D)\,\rho_{0n}^{(+)}+
iV_N\,\big(\rho_{00}^{(+)} -\sum_{n'=1}^N\rho_{n'n}^{(+)}\big)
-\gamma_-^{}\rho_{0n}^{(+)}+\gamma_+^{}\rho_{0n}^{(-)},
\label{a11e}\\
&\dot\rho_{0n}^{(-)}=i(E_n-E_0+D)\,\rho_{0n}^{(-)}+
iV_N\,\big(\rho_{00}^{(-)} -\sum_{n'=1}^N\rho_{n'n}^{(-)}\big)
-\gamma_+^{}\rho_{0n}^{(-)}+\gamma_-^{}\rho_{0n}^{(+)},
\label{a11f}
\end{eqnarray}
and $\rho_{n0}^{(\pm)}(t)=\rho_{0n}^{(\pm)*}(t),\ \rho_{n'n}^{(\pm)}(t)=\rho_{nn'}^{(\pm)*}(t)$.
Finally, one has to average over the noise, so that, $\rho_{jj'}^{}(t)=\rho_{jj'}^{(+)}(t)+\rho_{jj'}^{(-)}(t)$.
For more detailed arguments leading to Eqs.~(\ref{a11a})-(\ref{a11f}) and also for their exact microscopic quantum mechanical derivation for particular noise models, see Refs.~\cite{GM,Gurvitz1}.

\section{One-level acceptor}

\subsection{Steady state}

First, assume the acceptor is a one-level system, $N=1$. We introduce the variables, $\rho(t)=\rho_{}^{(+)}(t)+\rho_{}^{(-)}(t)$ and $\xi(t)=\rho_{}^{(+)}(t)-\rho_{}^{(-)}(t)$. Note, that $\rho_{00}^{}(t)+\rho_{11}^{}(t)=1$. In these variables, Eqs.~(\ref{a11a})-(\ref{a11f}) for $N=1$ are,
\begin{eqnarray}
&\dot\rho_{00}^{}=iV\big(\rho_{01}^{}
-\rho_{10}^{}\big),
\label{a12a}\\
&\dot\rho_{11}^{}=iV\big(\rho_{10}^{}
-\rho_{01}^{}\big),
\label{a12b}\\
&\dot\rho_{01}^{}=-i\epsilon\,\rho_{01}^{}+
iV\,\big(\rho_{00}^{}-\rho_{11}^{}\big)-iD\xi_{01}^{},
\label{a12c}\\
&\dot\xi_{00}^{}=iV\big(\xi_{01}^{}
-\xi_{10}^{}\big)
-2\gamma\,\xi_{00}^{}+2\eta\gamma\,\rho_{00}^{},
\label{a12d}\\
&\dot\xi_{11}^{}=iV\big(\xi_{10}^{}
-\xi_{01}^{}\big)
-2\gamma\,\xi_{11}^{}+2\eta\gamma\,\rho_{11}^{},
\label{a12e}\\
&\dot\xi_{01}^{}=-(i\epsilon+2\gamma)\xi_{01}^{}+
iV\big(\xi_{00}^{}-\xi_{11}^{}\big)
+(2\eta\gamma-iD)\rho_{01}^{},
\label{a12f}
\end{eqnarray}
where, $\gamma=(\gamma_+^{}+\gamma_-^{})/2$ and $\eta=(\gamma_+^{}-\gamma_-^{})/(\gamma_+^{}+\gamma_-^{})$.

Consider the density matrix in the asymptotic limit, $\rho (t\to\infty)\equiv \bar \rho$ and $\xi (t\to\infty)\equiv \bar \xi$.  If the density matrix reaches its steady-state in this limit, then $\dot{\bar\rho}=0$. In this case, Eqs.~({\ref{a12a})-({\ref{a12f}) can be easily solved. Indeed, it follows from (\ref{a12a}) that ${\rm Im}\,\bar\rho_{01}=0$. Substituting this into (\ref{a12c}), we find ${\rm Im}\,\bar\xi_{01}=0$. From the real parts of Eqs.~(\ref{a12d})-(\ref{a12f}) we find,
\begin{eqnarray}
&\bar\xi_{00}^{}=\eta\,\bar\rho_{00}^{},~~
\bar\xi_{11}^{}=\eta\,\bar\rho_{11}^{},~~
{\rm Re}\,\bar\xi_{01}^{}=\eta\,{\rm Re}\,\bar\rho_{01}^{}.
\label{re}
\end{eqnarray}
Taking the imaginary parts of Eqs.~(\ref{a12c}), (\ref{a12f}), we have,
\begin{eqnarray}
&-\epsilon {\rm Re}\,\bar\rho_{01}^{}+V\,(\bar\rho_{00}^{}-\bar\rho_{11}^{})
-D\,{\rm Re}\,\bar\xi_{01}^{}=0\nonumber,\\
&-\epsilon\, {\rm Re}\,\bar\xi_{01}^{}+V\,(\bar\xi_{00}^{}-\bar\xi_{11}^{})
-D\,{\rm Re}\,\bar\rho_{01}^{}=0.
\end{eqnarray}
Using Eqs.~(\ref{re}) one can rewrite these equations as,
\begin{eqnarray}
&-\epsilon {\rm Re}\,\bar\rho_{01}^{}+V\,(\bar\rho_{00}^{}-\bar\rho_{11}^{})
-\eta\,D\,{\rm Re}\,\bar\rho_{01}^{}=0\nonumber,\\
&-\epsilon\, {\rm Re}\,\bar\rho_{01}^{}+V\,(\bar\rho_{00}^{}-\bar\rho_{11}^{})
-{D\over\eta}\,{\rm Re}\,\bar\rho_{01}^{}=0.
\label{re11}
\end{eqnarray}
It immediately follows from these equations that, ${\rm Re}\,\bar\rho_{01}^{}={\rm Re}\,\bar\xi_{01}^{}=0$ and $\bar\rho_{00}^{}=\bar\rho_{11}^{}=1/2$. This implies equal distribution of the donor and acceptor in the asymptotic limit for any initial conditions. This result is drastically different from the no-noise case ($D=0$ or $\eta=\pm1$ in Eqs.~(\ref{re11})), considered in the previous section, where there is no steady-state, and the population of the acceptor at $t\to\infty$ remains very small if $\epsilon\gg V$. In the case of noise, however, the system always reaches equal distribution in the steady-state, no matter how small the noise is.

Note, that the equal distribution between the donor and the acceptor populations is always reached in the asymptotic limit, irrespectively of the relative position of the donor and the acceptor levels. This implies that the up and  down-hill transitions, generated by noise, proceed with the same probabilities. Usually such a behavior is considered as taking place in the high-temperature limit. However, this is not necessarily the case. Indeed, as we proved above, the equal populations of the donor and the acceptor in the asymptotic limit, takes place even for $\eta\not =0$ (see Eqs.~(\ref{re11})), corresponding to $p_+\not =p_-$. Thus, the equal population of the donor and the acceptor is related to the  non-equilibrium effect of noise. Indeed, noise is sustained in its steady state by an external source and therefore it is not affected by the  system. As a result, the average probabilities for the system of loosing and gaining energy from noise will be the same, resulting in the same occupation of the system states.

Now a natural question can be asked, what is a role of the system and noise parameters in the transition to equal distribution, since the later takes place for any values of these parameters. The answer is in the relaxation times which, for instance, can by very long, if noise is weakly coupled to the system.
The analysis of the relaxation times is a main subject of this paper. Now we are going to evaluate the transition rates by analyzing the time-dependent  Eqs.~(\ref{a12a})-(\ref{a12f}).}

\subsection{Transition time (rate)}

Consider, for simplicity, $\eta=0$. In order to solve Eqs.~({\ref{a12a})-({\ref{a12f}), we apply the Laplace transform, $\tilde\rho (E)=\int_0^\infty\rho(t)\,\exp (iEt)\,dt$, and correspondingly,
$\xi(t)\to\tilde\xi (E)$. Then,  Eqs.~({\ref{a12a})-({\ref{a12f}) become (see Ref.~\cite{GM,Gurvitz1}),
\begin{eqnarray}
&iE\trho_{00}^{}+iV\,\big(\trho_{01}^{}
-\trho_{10}^{}\big)=-i,
\label{a13a}\\
&iE\trho_{11}^{}+iV\,\big(\trho_{10}^{}
-\trho_{01}^{}\big)=0,
\label{a13b}\\
&(iE-i\epsilon)\trho_{01}^{}+
iV\,\big(\trho_{00}^{}-\trho_{11}^{}\big)-iD\txi_{01}^{}=0,
\label{a13c}\\
&(iE-2\gamma)\txi_{00}^{}+iV\big(\txi_{01}^{}
-\txi_{10}^{}\big)=0,
\label{a13d}\\
&(iE-2\gamma)\txi_{11}^{}+iV\big(\txi_{10}^{}
-\txi_{01}^{}\big)=0,
\label{a13e}\\
&(iE-i\epsilon-2\gamma)\txi_{01}^{}+
iV\big(\txi_{00}^{}-\txi_{11}^{}\big)
-iD\trho_{01}^{}=0,
\label{a13f}
\end{eqnarray}
and $\trho_{10}^{}(E)=\trho_{01}^{*}(-E)$.
Equations~({\ref{a13a})-({\ref{a13f})  can be rewritten in matrix form as,
\begin{eqnarray}
(i E\, {\mathbb I}  +M)\tilde R(E)=-R (0),
\end{eqnarray}
where the density matrix, $\tilde R(E)$ ($R (t)$), is written as an $8$-vector,
\begin{eqnarray}
\tilde R=\{\trho_{00}^{},\trho_{11}^{},
\trho_{01}^{},\trho_{10}^{},
\txi_{00}^{},\txi_{11}^{},
\txi_{01}^{},\txi_{10}^{}\}\, ,
\label{vec}
\end{eqnarray}
$M$ is an $8\times 8$ matrix corresponding to the r.h.s. part of Eqs.~({\ref{a13a})-({\ref{a13f}), and ${\mathbb I}$ is an $8\times 8$ unit matrix.
Solving these equations, we obtain rational expressions for the Laplace transformed density matrix elements,
\begin{eqnarray}
\tilde R_k(E)={\det [m_k(E)]\over\det [iE\,{\mathbb I}+M]},
\end{eqnarray}
where $k=\{1,2,\ldots ,8\}$, and $m_k$ is the corresponding minor determinant.

The density matrix as a function of $t$ is finally obtained via the inverse Laplace transform, Eq.~(\ref{invlap}),
\begin{eqnarray}
R_k(t)=\int\limits_{-\infty}^\infty \tilde R(E)\,
e^{-iEt}{dE\over 2\pi}.
\label{invl}
\end{eqnarray}
The secular and minor determinants can be represented by polynomials in powers of $E$. One finds for the secular determinant the following expression,
\begin{eqnarray}
\det [iE\, {\mathbb I}+M]=E(E+2i\gamma )\sum_{p=0}^6 A_pE^p\, ,
\label{secdet1}
\end{eqnarray}
where
\begin{eqnarray}
&A_0=-16D^2V^2\gamma^2\nonumber,\\
&A_1=2 i \gamma  \big[(D^2-\epsilon^2+4V^2)^2+16V^2\epsilon^2+4\gamma^2
(4V^2+\epsilon^2)\big],
\nonumber\\
&A_2=(D^2-\epsilon^2+4V^2)^2+16V^2\epsilon^2 +
4\gamma^2(12V^2+3\epsilon^2+2D^2),
\nonumber\\
&A_3=-8 i \gamma  \left(\gamma ^2+D^2+4 V^2+\epsilon ^2\right)\nonumber,\\
&A_4=-2 \left(6 \gamma ^2+D^2+4 V^2+\epsilon ^2\right),~~\\
&A_5=6i\gamma,~~A_6=1.
\label{coeff}
\end{eqnarray}

The integral in (\ref{invl}) can be calculated analytically by closing the integration contour in the lower half $E$-plane over the poles of denominator ($E=E_r$). One finds,
\begin{eqnarray}
\rho_k(t)=-i\sum_{r=1}^8{\det [m_k(E_r)]\over\prod_{r'\not =r} (E_r-E_{r'})}e^{-iU_{}^{(r)}t-\Gamma_{}^{(r)}t},
\end{eqnarray}
where $E_r=U_{}^{(r)}-i\Gamma_{}^{(r)}$ are the zeros of the secular determinant,
\begin{eqnarray}
\det [iE_r\, {\mathbb I}+M]=0,
\label{secdet2}
\end{eqnarray}
in the complex $E$-plane. The first pole, at $U_{}^{(1)}=\Gamma_{}^{(1)}=0$, produces a finite occupation in the asymptotic limit.
The second pole, at $U_{}^{(2)}=0$ and $\Gamma_{}^{(2)}=2\gamma$, produces a decay of the corresponding term with the rate, $2\gamma$. The remaining poles are obtained from the equation,
\begin{eqnarray}
A_0+A_1E+A_2E^2+\cdots +A_6E^6=0.
\label{poles}
\end{eqnarray}
The asymptotic transition rate, $\Gamma_1^{}$, is given by the pole with the smallest imaginary part,
$\Gamma_1^{}=\min\big\{\Gamma_{}^{(r)}\big\}$.

Now we will find an approximate analytical solution of Eq.~(\ref{poles}). Since we are looking for the pole with minimal value of the energy, we can keep only the first two terms in
Eq.~(\ref{poles}), neglecting the higher powers in $E$.
This yields the transition rate,  $\Gamma_1^{}={\rm Im}[A_0/A_1]$. Using Eq.~(\ref{coeff}), one can write explicitly for the transition time, $\tau_1^{}=1/\Gamma_1^{}$,
\begin{eqnarray}
\tau_1={(D^2-\epsilon^2-4V^2)^2+4\gamma^2\epsilon^2\over 8 \gamma  D^2 V^2}+{2(D^2+\gamma^2)\over \gamma D^2}.
\label{damp}
\end{eqnarray}
The accuracy of this procedure is determined by the parameter,
\begin{eqnarray}
\kappa = {A_0A_2\over A_1^2}={A_2\over 16 D^2V^2}\left({\Gamma_1^{}\over\gamma}\right)^2,
\label{accur}
\end{eqnarray}
which is expected to be less than one.

It follows from Eq.~(\ref{damp}), that the transition  time is minimal at $D^2=\epsilon^2+4V^2$ for  $\gamma\ll\epsilon$. This has a simple physical meaning: due to the influence of noise, the donor level fluctuates between $E_0\pm D$. This makes it partially in resonance with the acceptor' levels.  As a result, the rate of the donor-acceptor transitions increases. This effect is illustrated in Fig.~\ref{fig3}, where we show the occupation of the donor as a function of time for different values of the noise amplitude. The solid lines show $\rho_{00}^{}(t)$, obtained from the numerical solution of   Eqs.~({\ref{a12a})-({\ref{a12f}). The dashed lines are the asymptotic rate approximation,
\begin{eqnarray}
\rho_{00}^{}(t)={1\over2}\big(1+e^{-\Gamma_1^{}t}\big),
\label{damp1}
\end{eqnarray}
and $\Gamma_1^{}=1/\tau_1^{}$ is given by Eq.~(\ref{damp}). The following parameters are used (in arbitrary units): $\epsilon =5$, $V=1$, $\gamma=1$. The noise amplitudes are: $D=1$ (red lines), $D=5$ (blue lines), $D=10$ (black lines).
\begin{figure}[tbh]
\begin{center}
\scalebox{0.75}{\includegraphics{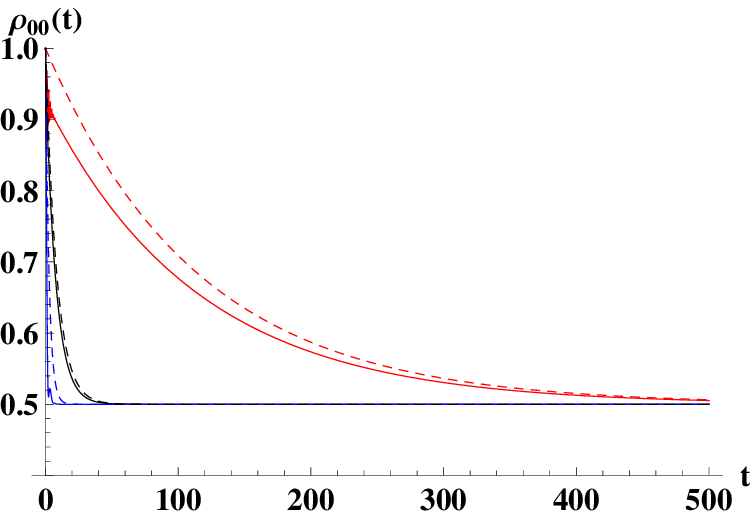}}
\end{center}
\caption{(Color online) Donor occupation a function of time, for $N=1$, $\eta=0$, $\epsilon =5$, $V=1$, $\gamma=1$, for 3 values of noise amplitude: $D=1$ (red lines), $D=5$ (blue lines) and $D=10$ (black lines). Solid lines show the results of the numerical solution of  Eqs.~(\ref{a12a})-(\ref{a12f}), and dashed lines correspond to the asymptotic approximation, Eq.~(\ref{damp1}).
\label{fig3}}
\end{figure}

It follows from this figure, that the asymptotic limit, given by Eq.~(\ref{damp1}), describes the behavior of $\rho_{00}^{}(t)$ quite well. As expected, the shortest transition time corresponds to, $D\simeq\epsilon$.

Now we consider an asymmetric noise, $\gamma_+\not =\gamma_-$ (or $\eta \not =0$), corresponding to different noise probabilities, $p_+\not =p_-$. Obviously, in the extreme case of very large asymmetry, $p_{+}=1$ and $p_{-}=0$  (or $\eta =1$), there is no noise effect on the donor-acceptor transition. Then, the donor (acceptor)  occupation stays very far from equal distribution, and does not reach the steady-state limit, Fig.~\ref{fig2}. However, for any other values of $\eta\not =1$, the probabilities of the donor and the acceptor occupations become equal and reach the steady state limit. Therefore, we anticipate very small transition rate when $\eta$ is very close to $1$.

This is illustrated in Fig.~\ref{fig3}, which shows the donor occupation as a function of time, obtained from the numerical solution of  Eqs.~(\ref{a12a})-(\ref{a12f}), for  $\epsilon =5$, $D=10$, $V=1$, $\gamma=1$, and three values of the noise asymmetry: $\eta=0$ (solid black line), $\eta =0.5$ (dashed blue line) and $\eta=0.9$ (dot-dashed red line), as a result of the numerical solution of  Eqs.~(\ref{a12a})-(\ref{a12f}).
\begin{figure}[tbh]
\begin{center}
\includegraphics{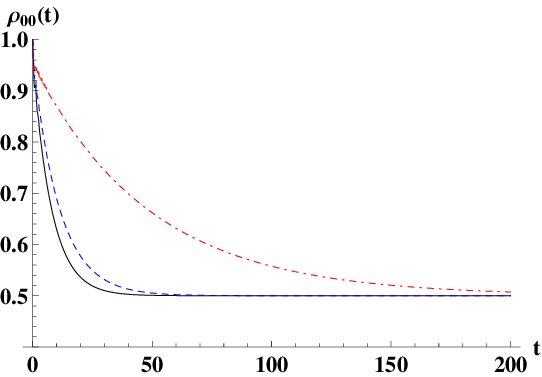}
\end{center}
\caption{(Color online) Donor occupation as a function of time for asymmetric noise, for $N=1$, $\epsilon =5$, $D=10$, $V=1$, $\gamma=1$, for three values of noise asymmetry: $\eta=0$ (solid black line), $\eta =0.5$ (dashed blue line) and $\eta=0.9$ (dot-dashed red line), as a result of the numerical solution of  Eqs.~(\ref{a12a})-(\ref{a12f}).
\label{fas1}}
\end{figure}
As expected, the donor reaches the asymptotic limit very slowly for $\eta =0.9$. However, otherwise it is not very different from the symmetric noise, $\eta =0$. Therefore, in the rest of this paper we concentrate only on $\eta =0$, mainly because the analytical formulas are most simple for interpretation, without loosing a generality.

\subsection{Reduced master equations}

Equations~(\ref{a12a})-(\ref{a12c}) resemble Bloch-type equations for the two-state density matrix, $\rho(t)$, except for the last term in Eq.~(\ref{a12c}), depending on $\xi_{01}(t)$. In fact, $\xi (t)$ is a function of $\rho (t)$, so that equations for $\rho (t)$ can be written in a closed form. This can be done by resolving Eqs.~(\ref{a13d})-(\ref{a13f}) for the Laplace transformed amplitudes, $\txi(E)$. Consider again, for simplicity, the case of $\gamma_+=\gamma_-=\gamma$. One finds,
\begin{eqnarray}
\txi_{01}^{}=D{[(E+2i\gamma)(E+2i\gamma+\epsilon)-2V^2]
\trho_{01}^{}+2V^2\trho_{10}^{}\over
(E+2i\gamma)[(E+2i\gamma)^2-\epsilon^2-4V^2]}.
\label{lap1}
\end{eqnarray}

The time-dependent amplitude, $\xi_{01}^{}(t)$, is obtained through the inverse Laplace transform (\ref{invlap}) of the amplitude, $\txi_{01}^{}(E)$, by closing the integration contour over the poles in the complex $E$-plane. Since we are interested in the asymptotic regime ($t\to\infty$), only the pole which is closest to zero survives. We therefore can replace $E\to 0$ in the prefactors of the amplitudes, $\trho$, in Eq.~(\ref{lap1}), thus obtaining,
\begin{eqnarray}
\xi_{01}^{}(t)=-iD{(2\gamma^2+V^2-i\gamma\epsilon)
\rho_{01}^{}(t)-V^2\rho_{10}(t)
\over \gamma(\epsilon^2+4V^2+4\gamma^2)}.
\end{eqnarray}
Substituting this result into Eq.~(\ref{a12c}), we obtain,
\begin{eqnarray}
&\dot\rho_{00}^{}=iV\big(\rho_{01}^{}
-\rho_{10}^{}\big),
\label{a14a}\\
&\dot\rho_{01}^{}=-i\epsilon'\rho_{01}^{}
+iV\,\big(2\rho_{00}^{}-1\big)
-\gamma_1^{}\rho_{01}^{}-\gamma_2^{}
(\rho_{01}^{}-\rho_{10}^{}),
\label{a14b}
\end{eqnarray}
where,
\begin{eqnarray}
\epsilon'=\epsilon\left(1-{D^2\over \epsilon^2+4V^2+4\gamma^2}\right),
\label{renen}
\end{eqnarray}
is a renormalized donor energy, and
\begin{eqnarray}
\gamma_1^{}={2\gamma D^2\over \epsilon^2+4V^2+4\gamma^2},~~~
\gamma_2^{}={V^2\over 2\gamma_{}^2}\,\gamma_1^{},
\label{blr}
\end{eqnarray}
are the damping rates.

Equations~(\ref{a14a})-(\ref{a14b}) have the form of Bloch equations for spin precession in a magnetic field in the presence of the environment. This can be seen by mapping the density matrix, $\rho(t)$, to the ``polarization'' vector, $\vec{S}(t)=\{S_x(t),S_y(t),S_z(t)\}$, via
$\rho(t)=[1+\vec{\tau}\cdot \vec{S}(t)]/2$, where $\tau_{x,y,z}$ are the Pauli matrices. Thus, we define
$S_z=2\rho_{00}^{}-1$, $S_y=i(\rho_{01}^{}-\rho_{10}^{})$ and $S_x=\rho_{01}^{}+\rho_{10}^{}$.
We find,
\begin{eqnarray}
&\dot S_z=2VS_y\nonumber,\\
&\dot S_y=-2VS_z-(\gamma_1+2\gamma_2)S_y+\epsilon'S_x \nonumber,\\
&\dot S_x=-\epsilon' S_y-\gamma_1S_x.
\label{bloch}
\end{eqnarray}
These equations coincide with the Bloch equations, where $\gamma_{1,2}$ are related to the two damping times, $T_1=\gamma_1^{-1}$ and $T_2=(\gamma_1^{}+2\gamma_2^{})^{-1}$.

Similar equations can be derived when the  interaction, $V$ (tunneling coupling) fluctuates, instead of fluctuating donor and acceptor energy levels \cite{GM,Gurvitz1}. However, the redox potential, $\epsilon$, would not be renormalized, as occurs in the  case of the energy-level fluctuations, Eq.~(\ref{renen}). This difference is essential for the electron transfer. Indeed, fluctuations of the energy levels can drive the donor and acceptor into resonance, which can greatly increase the transfer rate, Eq.~(\ref{damp}), Fig.~\ref{fig3}. In contrast, in the case of fluctuating coupling ($V$), resonance cannot occur, even though both of these noise-assisted processes are described by similar Bloch-type equations.

In the weak interaction limit, $V\ll D,\gamma$, one finds that $T_1=T_2$. Then, Eqs.~(\ref{a14a})-(\ref{a14b}) become further simplified. Solving these equations in this limit, we find for the asymptotic  transition rate,
\begin{eqnarray}
\Gamma_1^{}=\frac{8 \gamma  D^2 V^2}
{(D^2-\epsilon^2)^2+4\gamma^2
\epsilon^2},
\label{damp2}
\end{eqnarray}
which coincides with Eq.~(\ref{damp}) in the same limit.

The weak interaction limit can be very useful for the multi-level case, since it greatly simplifies the treatment, without losing any physical features of the process.
We show in Fig.~\ref{fig4} the occupation of donor, $\rho_{00}(t)$, for different values of the noise amplitude. The solid lines correspond to Eqs.~(\ref{a12a})-(\ref{a12f}), whereas the dashed lines are obtained from the reduced master equations~(\ref{bloch}) in the limit of weak interaction: $\gamma_2^{}=0$ and $\gamma_1^{}=2\gamma D^2/(\epsilon^2+4\gamma_{}^2)$. The parameters are the same as in Fig.~\ref{fig3}.
\begin{figure}[tbh]
\begin{center}
\scalebox{0.75}{\includegraphics{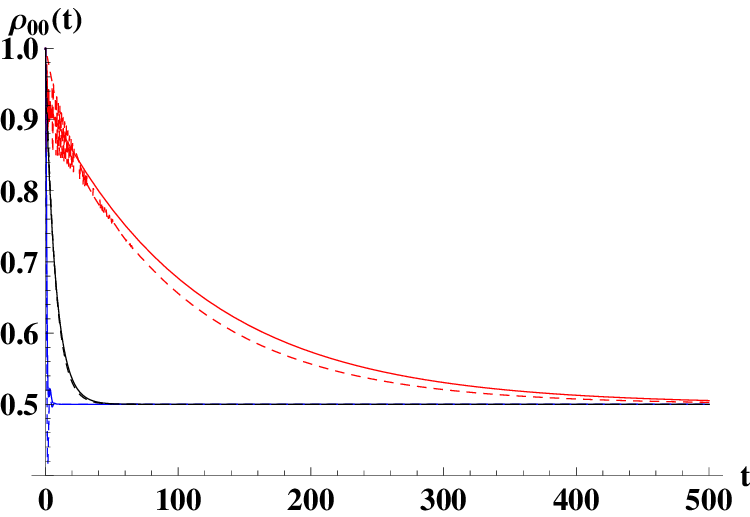}}
\end{center}
\caption{(Color online) Donor occupation as a function of time, for $N=1$, for parameters the same as in Fig.~\ref{fig3}. Solid lines show the result of the numerical solution of Eqs.~(\ref{a12a})-(\ref{a12f}), and dashed lines correspond to Eqs.~(\ref{a14a})-(\ref{a14b}) in the limit of small $V$.
\label{fig4}}
\end{figure}
It follows from this figure that the reduced Bloch-type master equations describe  the asymptotic limit very well, even for $V\sim \gamma$.

\section{$N$-level acceptor}

Consider Eqs.~(\ref{a11a})-(\ref{a11f}). As in the previous section, we rewrite these equations in the variables, $\rho=\rho_{}^{(+)}+\rho_{}^{(-)}$
and $\xi=\bar\rho_{}^{(+)}-\rho_{}^{(-)}$. Then, these equations can be rewritten as,
\begin{eqnarray}
&\dot\rho_{00}^{}=iV_N\sum_{n=1}^N(\rho_{0n}^{}-\rho_{n0}^{}),
\label{a15a}\\
&\dot\rho_{nn'}^{}=-i\epsilon_{nn'}^{} \,\rho_{nn'}^{}+
iV_N\,\big(\rho_{n0}^{}
-\rho_{0n'}^{}\big),
\label{a15b}\\
&\dot\rho_{0n}^{}=
i\epsilon_{n0}^{}\,\rho_{0n}^{}+
iV_N\,\Big(\rho_{00}^{}
-\sum_{n'=1}^N\rho_{n'n}^{}\Big)
-iD\,\xi_{0n}^{},
\label{a15c}\\
&\dot\xi_{00}^{}=-2\gamma\,\xi_{00}^{}+
iV_N\,\sum_{n=1}^N(\xi_{0n}^{}-\xi_{n0}^{})
-2\eta\gamma\,\rho_{00}^{},
\label{a15d}\\
&\dot\xi_{nn'}^{}=(i\epsilon_{n'n}^{}
-2\gamma)\,\xi_{nn'}^{}
+iV_N\big(\xi_{n0}^{}-\xi_{0n'}^{}\big)
+2\eta\gamma\,\rho_{nn'}^{},
\label{a15e}\\
&\dot\xi_{0n}^{}=(i\epsilon_{n0}^{}-2\gamma)\,\xi_{0n}^{}+
iV_N\,\Big(\xi_{00}^{}
-\sum_{n'=1}^N\xi_{n'n}^{}\Big)  +(2\eta\gamma-iD)\,\rho_{0n}^{},
\label{a15f}
\end{eqnarray}
where $\epsilon_{n'n}=E_{n'}-E_n$.

\subsection{Degenerate case}

Consider the case when all energy levels of the acceptor coincide: $\epsilon_{nn'}^{}=0$ for $n,n'=1,\ldots ,N$. Then, one can sum over the acceptor states, $n$, in Eqs.~(\ref{a15a})-(\ref{a15f}), thus obtaining,
\begin{eqnarray}
&\dot\rho_{00}^{}=iV(\varrho_{01}^{}-\varrho_{10}^{}),
\label{a16a}\\
&\dot\varrho_{11}^{}=
iV\big(\varrho_{10}^{}
-\varrho_{01}^{}\big),
\label{a16b}\\
&\dot\varrho_{01}^{}=
-i\epsilon\,\varrho_{01}^{}+
iV\Big(\rho_{00}^{}
-\varrho_{11}^{}\Big)
-iD\,\zeta_{01}^{},
\label{a16c}\\
&\dot\xi_{00}^{}=
iV(\zeta_{01}^{}-\zeta_{10}^{})
-2\gamma\,\xi_{00}^{}
-2\eta\gamma\,\rho_{00}^{},
\label{a16d}\\
&\dot\zeta_{11}^{}=-2\gamma\,\zeta_{11}^{}
+iV\big(\zeta_{10}^{}-\zeta_{01}^{}\big)
+2\eta\gamma\,\varrho_{11}^{},
\label{a16e}\\
&\dot\zeta_{01}^{}=-(i\epsilon+2\gamma)\,\zeta_{01}^{}+
iV\Big(\xi_{00}^{}
-\zeta_{11}^{}\Big)
+(2\eta\gamma-iD)\,\varrho_{01}^{},
\label{a16f}
\end{eqnarray}
where
\begin{eqnarray}
&\varrho_{11}^{} = {1\over N}\sum^N_{n,n'=1}\rho_{nn'}^{}, \quad \varrho_{01}^{} = {1\over \sqrt{N}}\sum^N_{ n=1}\rho_{0n}^{}, \\
&\zeta_{11}^{} ={1\over N} \sum^N_{n,n'=1}\xi_{nn'}^{},\quad
\zeta_{01}^{} = {1\over \sqrt{N}}\sum^N_{ n=1}\xi_{0n}^{}.
\end{eqnarray}

One finds that Eqs.~(\ref{a16a})-(\ref{a16f}) coincide with Eqs.~(\ref{a12a})-(\ref{a12f}), describing time-evolution of the density matrix, $\rho (t)$, for the case of one-level acceptor. As follows from these equations, $\rho_{00}^{}(t)+\varrho_{11}^{}(t)=$const. However, in contrast with the one-level acceptor, this constant is not unity, in general, and its value depends on the initial conditions. Indeed, $\varrho_{11}^{}(t)$ is not the occupation probability of acceptor, since it includes the off-diagonal density matrix elements. It can be easily seen by rewriting it explicitly,
\begin{eqnarray}
\varrho_{11}^{}(t)={1\over N}\sum_{n=1}^N\rho_{nn}^{}(t)+{1\over N}\sum_{n\not =n'}^N{\rm Re}\,\rho_{nn'}^{}(t).
\end{eqnarray}

Consider now the asymptotic limit, where we denote,  $\bar\rho_{00}^{}=\rho_{00}^{}(t\to\infty )$ and $\bar\varrho_{11}^{}=\varrho_{11}^{}(t\to\infty )$.
Using the same transformations as in Sec. IV A, we arrive at $\bar\rho_{00}^{}=\bar\varrho_{11}^{}$, Eqs.~(\ref{re11}). Thus, the probability of finding the donor occupied in the steady-state is,
\begin{eqnarray}
\bar\rho_{00}^{}=&{1\over2}\big[\rho_{00}^{}(0)
+\varrho_{11}^{}(0)\big] ={1+(N-1)\rho_{00}(0)\over 2N} +{1\over
2N}\sum_{n\not =n'}^N{\rm Re}\,\rho_{nn'}^{}(0),
\label{degen}
\end{eqnarray}
where we used the normalization condition, $\rho_{00}^{}(0)+\sum_{n=1}^N\rho_{nn}^{}(0)=1$.

In contrast with the case of one-state acceptor, the off-diagonal density matrix elements (coherences) do not vanish in the steady state limit. One finds,
\begin{eqnarray}
\sum_{n\not =n'}^N{\rm Re}\,\bar\rho_{nn'}^{}=(N+1)\bar\rho_{00}^{}-1.
\label{degen1}
\end{eqnarray}
This is the case of partial decoherence \cite{Gurvitz1,amnon}, which takes place when the quantum system possesses a symmetry, that cannot be destroyed by the environment.

It follows from Eq.~(\ref{degen}), that the steady-state distribution depends
on the initial state. In particular, $\bar\rho_{00}^{}=1/2$ for the initially
occupied donor, $\rho_{00}^{}(0)=1$. However,
from the previous study of the electron transport in a similar system
\cite{gur1}, one would expect equal occupation of {\em all}  levels in the
limit of $t\to\infty$, for any value of the noise amplitude and  level spacing.
This would be drastically different from the degenerate level case. To
understand this problem, we investigate the non-degenerate case in detail.

\subsection{Asymptotic state}

Consider now the asymptotic limit, $\rho(t\to\infty)=\bar\rho$, and $\xi(t\to\infty)=\bar\xi$, where the acceptors energy levels  are not degenerate. Since in this limit $\dot\rho(t)\to 0$ and $\dot\xi(t)\to 0$, we can rewrite Eqs.~(\ref{a15a})-(\ref{a15f}) as,
\begin{eqnarray}
&\rho_{00}^{}+\sum_{n=1}^N\rho_{nn}=1,
\label{a17a}\\
&\epsilon_{n'n}^{}\,\rho_{nn'}^{}
+V_N\,\big(\rho_{n0}^{}
-\rho_{0n'}^{}\big)=0,
\label{a17b}\\
&\epsilon_{n0}^{}\,\rho_{0n}^{}+
V_N\,\Big(\rho_{00}^{}
-\sum_{n'=1}^N\rho_{n'n}^{}\Big)
-D\,\xi_{0n}^{}=0,
\label{a17c}
\end{eqnarray}
\begin{eqnarray}
&2V_N\,\sum_{n=1}^N{\rm Im}\,\xi_{0n}^{}
+2\gamma\,\xi_{00}^{}=0,
\label{a17d}\\
&\big(i\epsilon_{n'n}^{}-2\gamma\,\big)\xi_{nn'}^{}
+iV_N\,\big(\xi_{n0}^{}-\xi_{0n'}^{}\big)=0,
\label{a17e}\\
&(i\epsilon_{n0}^{}-2\gamma )\,\xi_{0n}^{}+
iV_N\,\Big(\xi_{00}^{}
-\sum_{n'=1}^N\xi_{n'n}^{}\Big)
-iD\rho_{0n}^{}=0,
\label{a17f}
\end{eqnarray}
where $\epsilon_{jj'}^{}=E_j-E_{j'}$. For simplicity, we assume $\gamma_+^{}=\gamma_-^{}=\gamma$.

Since $\epsilon_{nn'}^{}\not =0$, we obtain from Eq.~(\ref{a17b})-(\ref{a17d}),
\begin{eqnarray}
{\rm Im}\,\rho_{0n}^{}={\rm Im}\,\rho_{nn'}^{}={\rm Im}\,\xi_{0n}^{}=\xi_{00}^{}=0.
\end{eqnarray}
Then, one finds from Eqs.~(\ref{a17e}) and (\ref{a17f}),
\begin{eqnarray}
&\xi_{n'n}^{}=iV_N{{\rm Re}\,(\xi_{0n}-\xi_{n'0})\over i\epsilon_{nn'}^{}-2\gamma}\label{d1a},\\
&(i\epsilon_{n0}^{}-2\gamma )\,{\rm Re}\,\xi_{0n}^{}+
V_N^2\,
\sum_{n'=1}^N{{\rm Re}\,(\xi_{0n}-\xi_{n'0})\over i\epsilon_{nn'}^{}-2\gamma}=iD{\rm Re}\,\rho_{0n}^{}.
\label{d1b}
\end{eqnarray}
Taking the real part of Eq.~(\ref{d1b}) we find,
\begin{eqnarray}
X_n- \sum_{n'}\frac{C_{nn'}}{1+\sum_{n'}C_{nn'}}X_{n'}=0,
\label{d2}
\end{eqnarray}
where $X_n={\rm Re}\,\xi_{0n}^{}$ and $C_{nn'}=V_N^2/(\epsilon_{nn'}^2+4\gamma^2)$. This is a system of coupled linear equations, $\sum_{n'=1}^N{\cal A}_{nn'}^{(N)}X_{n'}=0$, where $\det {\cal A}_{}^{(N)}>0$. (See Appendix A for details.) As a result, $X_n={\rm Re}\,\xi_{0n}^{}=0$, for all $n$. Inserting this into Eq.~(\ref{d1b}), we find that ${\rm Re}\,\rho_{0n}^{}=0$. Substituting this result into Eq.~(\ref{a17b}), one finds ${\rm Re}\,\rho_{nn'}^{}=0$ for $n\not = n'$. Then, using Eqs.~(\ref{a17a}) and (\ref{a17c}), we find equal occupation of all levels in the asymptotic limit of $t\to\infty$,
\begin{eqnarray}
\rho_{00}^{}=\rho_{nn}^{}={1\over N+1}.
\label{asocc}
\end{eqnarray}
This implies that for $N\gg 1$, the probability of finding the acceptor occupied is close to unity. So, the efficiency of the electron (exciton) transfer can be close to 100\%.

It is remarkable that the above result has been derived for any value of spacing between the acceptor' levels  ($\epsilon_{nn'}$), no matter how small. However,  for the case of the exact degenerate acceptor' levels, $\epsilon_{nn'}=0$,
this proof is not valid, since we cannot use Eq.~(\ref{a17b}) to obtain ${\rm Re}\,\rho_{nn'}^{}=0$ for $n\not = n'$. Indeed, this quantity is not zero in the degenerate case, as follows from  Eq.~(\ref{degen1}). Moreover, the asymptotic distribution depends on the initial conditions, Eq.~(\ref{degen}). One finds, $\bar\rho_{00}^{}=1/2$, for the initial condition corresponding to the occupied donor, instead of $1/(N+1)$ for the non-degenerate case, no matter how small the acceptor' band-width, $\delta_a$, Eq.~(\ref{acen}).

Thus the above result displays discontinuity of the steady-state occupation with the acceptor's bandwidth, $\delta_a$. The only possible explanation of this discontinuity can be found in an analysis of the donor-acceptor transfer dynamics. We can anticipate that the system reaches first its equal donor-acceptor occupation ($1/2$). Then, it is distributed inside the acceptor's states, finally approaching equal occupation for all levels, $1/(N+1)$. One expects that the second transition rate would depend on the level spacing, and that decreases to zero when the acceptor's levels are degenerate. In this case, the equal distribution, $1/(N+1)$, can never be reached, so that the system stays asymptotically in the equal donor-acceptor occupation $(1/2)$. Such an explanation can resolve the discontinuity problem, and it will be confirmed by the following  analysis of the transfer dynamics.

\subsection{Reduced master equations}

In order to determine the transition rate, we first reduce Eqs.~(\ref{a15a})-(\ref{a15f}) to simplified equations that involve only  the density matrix, $\rho (t)$. This is similar to the case of Sec. 5.3, for $N=1$. For this reason, we will express $\xi (t)$ in terms of $\rho(t)$ using Eqs.~(\ref{a15d})-(\ref{a15f}). For simplicity, we assume $\eta =0$, and consider the limit of $V_N\ll \gamma, D$. Then, we can neglect the term proportional to $V_N$ in Eq.~(\ref{a15f}), thus obtaining,
\begin{eqnarray}
\xi_{0n}(t)=-{iD\over i\epsilon_{0n}^{}+2\gamma}\,\rho_{0n}(t).
\end{eqnarray}
Substituting this result in Eq.~(\ref{a15c}), we find,
\begin{eqnarray}
&\dot\rho_{00}^{}=iV_N\sum_{n=1}(\rho_{0n}^{}-\rho_{n0}^{}),
\label{bloch2a}\\
&\dot\rho_{nn'}^{}=-i\epsilon_{nn'}^{}\,\rho_{nn'}^{}+
iV_N\,\big(\rho_{n0}^{}
-\rho_{0n'}^{}\big),
\label{bloch2b}\\
&\dot\rho_{0n}^{}=
-(i\epsilon'_{0n}+\gamma_{0n}^{})\,\rho_{0n}^{}+
iV_N\,\Big(\rho_{00}^{}
-\sum_{n'=1}^N\rho_{n'n}^{}\Big),
\label{bloch2c}
\end{eqnarray}
where $V_N=V/\sqrt{N}$ and,
\begin{eqnarray}
\gamma_{0n}^{}={2\gamma D^2\over \epsilon_{0n}^2+4\gamma^2_{}}~~{\rm and}~~
\epsilon_{0n}^{\prime}=\epsilon_{0n}^{}\left(1-{D^2
\over \epsilon_{0n}^2+4\gamma^2_{} }\right).
\label{bloch3}
\end{eqnarray}
These equations are similar to the Bloch-type equations (\ref{bloch}) for a two-state system, but now they have been extended to include a multi-level acceptor. The most pronounced feature of these equations is the renormalization  of the donor-acceptor energies, $\epsilon_{0n}^{}\to \epsilon'_{0n}$ (Eq.~(\ref{bloch3})), due to noise. This reconstruction of the redox potential is given by: $\Delta \epsilon_{0n}/\epsilon_{0n}=-D^2/(\epsilon^2_{0n}+4\gamma^2)$. This is analogous to the reconstruction energy in the Marcus ET rate, but for a multi-level acceptor, and a noisy (instead of thermal) environment \cite{HDR,MRSN}. As a result, the transition rate can be greatly increased if $\epsilon'_{0n}\simeq 0$, when the donor and  acceptor are effectively in resonance.

We show in Fig.~\ref{fig5} the donor occupation as a function of time, obtained from Eqs.~(\ref{a15a})-(\ref{a15f}) (black lines), in comparison with the same quantity obtained from the reduced master equations~(\ref{bloch}) (red lines) for $N=2$ and $N=10$. The parameters are: $\epsilon =5$, $V=1$, $\gamma=1$, and $D=5$,  corresponding to the effective ``resonance conditions". The acceptor levels, $E_n$, are given by Eq.~(\ref{acen}), where the acceptor bandwidth, $\delta_a=1$, is the same for $N=2$ and $N=10$. The dashed lines show the asymptotic limit, $1/(N+1)$. The results shown in Fig.~\ref{fig5} clearly demonstrate that the reduced master equations are a very good approximation, even beyond  the condition, $V_N\ll \gamma,D$, used for their derivation.
\begin{figure}[tbh]
\begin{center}
\scalebox{0.75}{\includegraphics{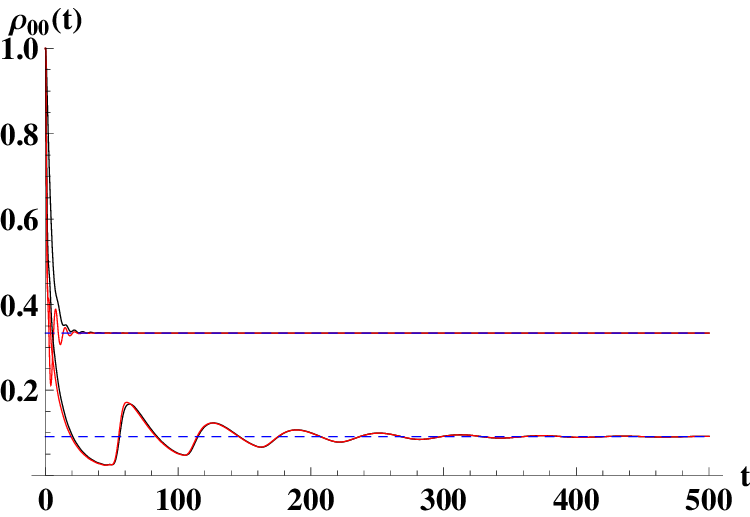}}
\end{center}
\caption{(Color online) Donor occupation for the two cases of acceptor levels, $N=2,10$, but with the same bandwidth, $\delta_a=1$. The noise amplitude, $D=\epsilon =5$.
The other parameters are the same as in Fig~\ref{fig3}. The dashed lines correspond to the asymptotic limit, $1/(N+1)$.
\label{fig5}}
\end{figure}

It follows from Fig.~\ref{fig5}, that the characteristic time, $\tau_N$, needed to approach equal occupation of all states of the system is much longer for $N=10$, than for $N=2$, although the bandwidth, $\delta_a$, is the same for  both cases. (See also \cite{MB}, where the case $N=2$ was considered for a thermal environment.) This dependence on $N$ is not trivial, and at first glance, even counter-intuitive. Indeed, if the bandwidth, $\delta_a$, of the acceptor is fixed and $N$ increases, then the acceptor density of states increases. So, intuition tells us that the ET rate may increase. According to Fig.~\ref{fig5}, it indeed increases, but only for the initial stage of the acceptor' population. At later times, other factors become important.   For the parameters chosen in Fig.~\ref{fig5}, the bandwidth, $\delta_a$ is 5 times smaller than $\epsilon$. In this case, two characteristic regimes can be expected. The first one is a relatively rapid population of the acceptor with the population close to 1/2, as for two-level system. The ET rate for this stage is greater for $N=10$ compared with $N=2$. The second regime involves a re-population of the acceptor levels. This requires additional time, and is accompanied by oscillations which are clearly seen in Fig.~\ref{fig5}. The period of these oscillations can be estimated to be: $T=2\pi N/\delta$. So, even if the asymptotic efficiency (probability) of the acceptor' population, $P_a(t\rightarrow\infty)=N/N+1$, increases with $N$ increasing, the time of approaching this asymptotic regime can increase with $N$. Below, we analyze this dependence on both $N$ and $\delta$, and show that the following scaling exists: $\tau_N\propto(N/\delta)^2$, for large $N$. These results are very important for engineering of this type of bio-complexes in order to achieve optimal ET rates and efficiencies. Below, we derive
analytical formulas for the transition time as a function of level spacing.

\subsection{Transition rates}

We now evaluate the transfer rates, using Eqs.~(\ref{bloch2a})-(\ref{bloch2c}). We start with $N=2$. Then, the Laplace transform of these equations can be written:
\begin{eqnarray}
&iE\trho_{00}^{}+iV_2\,\big(\trho_{01}^{}
-\trho_{10}^{}\big)+iV_2\,\big(\trho_{02}^{}
-\trho_{20}^{}\big)=-i,
\label{d4a}\\
&iE\trho_{11}^{}+iV_2\,\big(\trho_{10}^{}
-\trho_{01}^{}\big)=0,
\label{d4b}\\
&iE\trho_{22}^{}+iV_2\,\big(\trho_{20}^{}
-\trho_{02}^{}\big)=0,
\label{d4c}\\
&(iE-i\delta)\trho_{12}^{}+iV_2\,\big(\trho_{10}^{}
-\trho_{02}^{}\big)=0,
\label{d4d}\\
&\Big(iE-i\epsilon'_{01}-\gamma_{01}^{}\Big)\trho_{01}^{}+
iV_2\,\big(\trho_{00}^{}-\trho_{11}^{}-\trho_{21}^{}\big)
=0,
\label{d4e}\\
&\Big(iE-i\epsilon'_{02}-\gamma_{02}^{}\Big)\trho_{02}^{}+
iV_2\,\big(\trho_{00}^{}-\trho_{22}^{}-\trho_{12}^{}\big),
=0,
\label{d4f}
\end{eqnarray}
where $\delta=\epsilon_{12}^{}$, $V_2=V/\sqrt{2}$, and $\epsilon'_{01(2)}$, $\gamma_{01(2)}^{}$ are given by Eq.~(\ref{bloch3}), with $\epsilon_{01(2)}=\epsilon\mp{\delta/2}$.

The asymptotic transition rate, $\Gamma_2^{}$, is obtained from Eq.~(\ref{secdet2}) by expanding the secular determinant in powers of $E$, Eq.~(\ref{poles}), and keeping only the lowest power term, which dominates the asymptotic behavior. By neglecting  the higher order terms proportional to $\delta^4$ and $V^2\delta^2$, we obtain for the transition time $\tau_2^{}=1/\Gamma_2^{}$,
\begin{eqnarray}
\tau_2^{}={16\over 3}\tau_1^{}+{2V^2\,(4\gamma^2+\epsilon^2)\over 3\,\gamma\, D^2\,\delta^2},
\label{asrate}
\end{eqnarray}
where $\tau_1^{}$ is the transition time for $N=1$, Eq.~(\ref{damp}).

It follows from Eq.~(\ref{asrate}) that in the limit of $\delta\to 0$, the time required for equal occupation of all acceptor' states, Eq.~(\ref{asocc}), diverges as $1/\delta^2$. This is not surprising, since this limit corresponds to the degenerate case, considered in Sec. 5.1. Then, all levels of the acceptor effectively become a single level. As a result, the acceptor is only partially occupied in the asymptotic limit. Indeed, its occupation approaches $1/2$, with the corresponding occupation rate $\Gamma_1$, Eq.~(\ref{damp2}). This implies that for $\delta\not =0$, the acceptor occupation (depletion of donor) can be described effectively by two rates. The first, $\Gamma_1^{}$, populates the acceptor up to an occupation of $1/2$. Then, the process proceeds with the second rate, $\Gamma_2^{}\ll \Gamma_1^{}$, to the asymptotic population of $2/3$.
Neglecting the first term in Eq.~(\ref{asrate}), we can write,
\begin{eqnarray}
\Gamma_2^{}={3\,\gamma\, D^2\,\delta^2\over 2V^2\,(4\gamma^2+\epsilon^2)}.
\label{asocc3}
\end{eqnarray}
In fact, both transition rates, given by Eqs.~(\ref{damp2}) and (\ref{asocc3}), can be obtained from the same Eq.~(\ref{secdet2}), by keeping the higher order terms in the expansion of the secular determinant in powers of $E$.
\begin{figure}[tbh]
\begin{center}
\scalebox{0.75}{\includegraphics{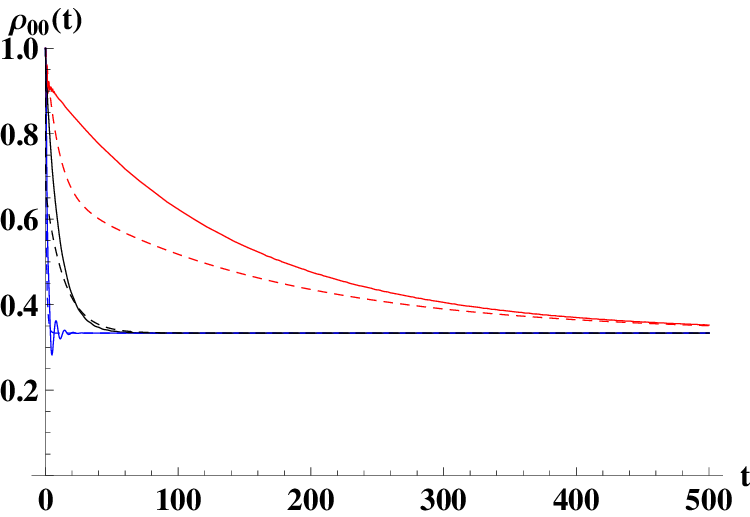}}
\end{center}
\caption{(Color online) Donor occupation as a function of time, for $N=2$,
$\epsilon =5$, $V=1$, $\gamma=1$, $\delta_a =1$, for 3 values of the
noise amplitude: $D=1$ (red lines), $D=5$ (blue lines) and $D=10$ (black
lines). Solid lines show the results of the numerical solution of
Eqs.~(\ref{a15a})-(\ref{a15f}), and the dashed lines correspond to
Eq.~(\ref{damp4}).
\label{fig6}}
\end{figure}

As in Eq.~(\ref{damp1}), but now with two exponents,  one can effectively represent the donor-acceptor transitions as,
\begin{eqnarray}
\rho_{00}(t)={1\over6}(1+e^{-\Gamma_1^{} t})(2+e^{-\Gamma_2^{} t}).
\label{damp4}
\end{eqnarray}
This simple formula describes the dynamics of the donor-acceptor transition reasonably well, as demonstrated in Fig.~\ref{fig6}. We display there the donor occupation, given by Eq.~(\ref{damp4}) (dashed lines), for different values of the noise amplitude, $D$, in comparison with the exact calculations, Eqs.~(\ref{a15a})-(\ref{a15f}), shown by the solid lines.

It is remarkable that the second rate, $\Gamma_2^{}$, is not of the
Breit-Wigner type, in contrast with $\Gamma_1^{}$, Eq.~(\ref{damp2}). This
supports the two-step dynamics of the donor-acceptor transitions. Indeed,
during the first step the transferred electron does not ``discern'' individual
levels of the acceptor. Therefore, one can anticipate that the corresponding
rate, $\Gamma_1^{}$, is independent of the
number of acceptor states ($N$) and has a Lorentzian-type shape as a function of the noise amplitude.

At the next step, the electron is redistributed among the acceptor' states. Since the energy spread of these states is narrow, there is no reason to expect any Lorentzian-type dependence of the second rate $\Gamma_2^{}$ on the noise amplitude, $D$. It it natural to expect the second rate $\Gamma_2^{}$ would always increase with $D$, as it is demonstrated by Eq.~(\ref{asocc3}). This point could be very important, since a large noise amplitude can compensate the decrease of $\Gamma_2^{}$ with a decrease of the level splitting.

The proposed two-step dynamics for  ET suggests a natural extension of our
results to any number of acceptor states, $N$. This can be done by replacing
the bandwidth, $\delta_a$, in Eq.~(\ref{asocc3}) for $N=2$ by the ratio,
$\delta /(N-1)$, for any $N\geq 2$, which represents the level spacing
(inverse density of states). Then, the corresponding transition time for
occupation of all acceptor states $(N)$ would be
$\tau_N^{}=1/\Gamma_N^{}$. The partial occupation of $n<N$ acceptor
states requires less time, of course. The transition time can be evaluated
using the same expression, obtained from Eqs.~(\ref{damp2}), (\ref{asrate}),
by replacing $\delta_a$ by $\delta_a/(n-1)$. Finally, the total transition time
becomes,
\begin{eqnarray}
\tau_n^{}\propto \frac{(D^2-\epsilon^2)^2+4\gamma^2
\epsilon^2}{\gamma  D^2 V^2}+
{V^2\,(4\gamma^2+\epsilon^2)\over
\gamma\, D^2\,\delta_a^2}(n-1)^2.
\label{asocc4}
\end{eqnarray}
Here the first term is a ``coarse-grained'' time, $\propto 1/\Gamma_1$,
Eq.~(\ref{damp2}), when the acceptor can be considered as a single level.
The maximal acceptor occupation in this case can reach only 1/2.
Subsequently,  the electron is redistributed among the acceptor states. The
corresponding time is given by the second term in Eq.~(\ref{asocc4}).
\footnote{Note,  that the coarse-grained dynamics, described by Eq.~(\ref{asocc4}), cannot be valid for a ``Markovian'' acceptor (with infinite band-width and constant density of state), which generates a pure exponential decay, (see Sec. 2). Formally it follows from Eq.~(\ref{asrate}), obtained as an expansion in powers of $\delta^2/(V^2+\epsilon^2)$.}

Interesting ``scaling" regime occurs when the noise amplitude, $D= \epsilon$, and $\gamma \ll \epsilon$. In this case,  the transition time is given by,
\begin{eqnarray}
\tau_N^{}\propto {4\gamma\over V^2}+{V^2\over\gamma}\Big({N-1\over\delta_a}\Big)^2.
\label{scaling}
\end{eqnarray}
It is quite remarkable that in this case the transition time does not depend
on the redox potential, $\epsilon$, and on the noise amplitude, $D$, but
only on its spectral width, $\gamma$. Moreover, it scales with $\gamma
/V^2$.

This prediction can be verified by a direct evaluation of the donor occupation, $\rho_{00}^{}(t)$, for different values of $\gamma$, by keeping $V$, and under the condition, $D=\epsilon$. The results of the calculations, using the exact equations, Eqs.~(\ref{a15a}-\ref{a15f}), are shown in Fig.~\ref{fig7}, for $N=20$.
\begin{figure}[tbh]
\begin{center}
\scalebox{0.55}{\includegraphics{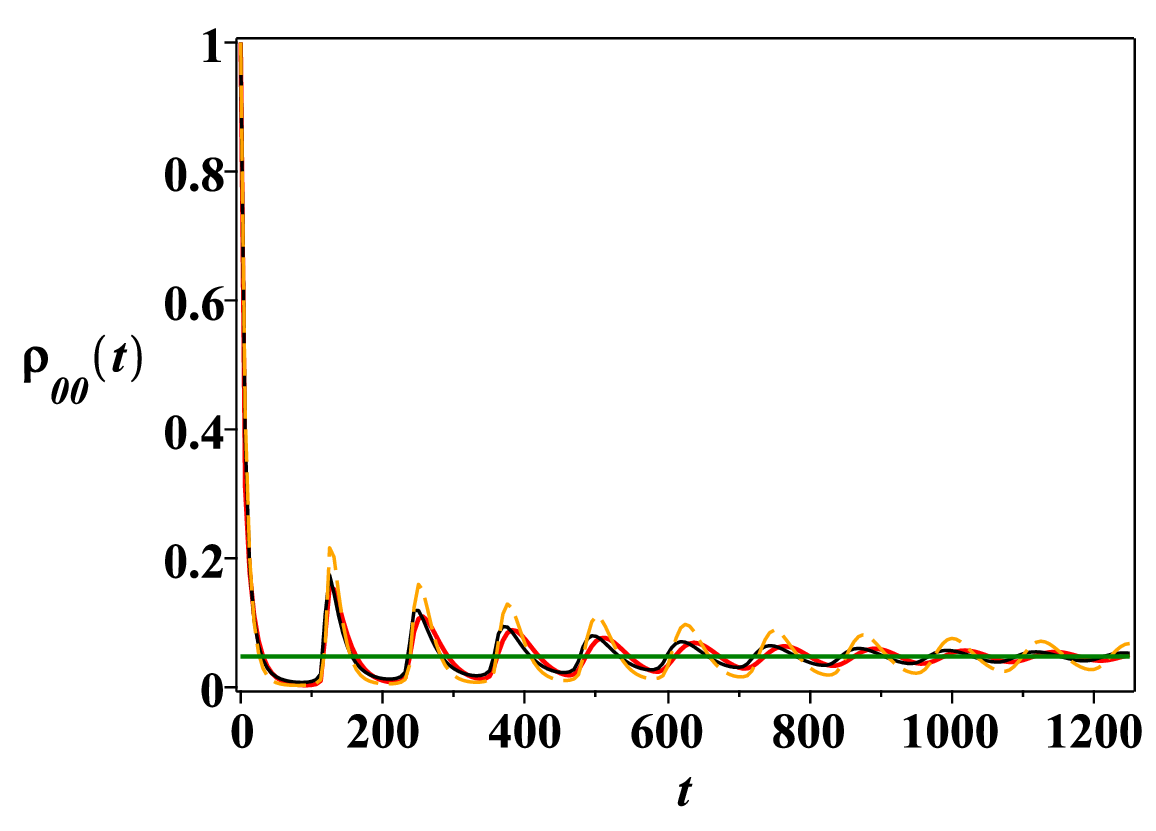}}
\end{center}
\caption{(Color online) Donor occupation as a function of time, for $N=20$: $\epsilon =D=5$ and
$\delta_a =1$. Red and black curves correspond to
$V=1$, $\gamma=1$ and $V=0.5$, $\gamma =0.25$, respectively.
Dashed-orange curve corresponds to values of $V$ and $\gamma$ out of
the scaling: $V=0.5$, $\gamma =1$. The asymptotic limit is shown by green line.
\label{fig7}}
\end{figure}

\begin{figure}[tbh]
\begin{center}
\scalebox{0.55}{\includegraphics{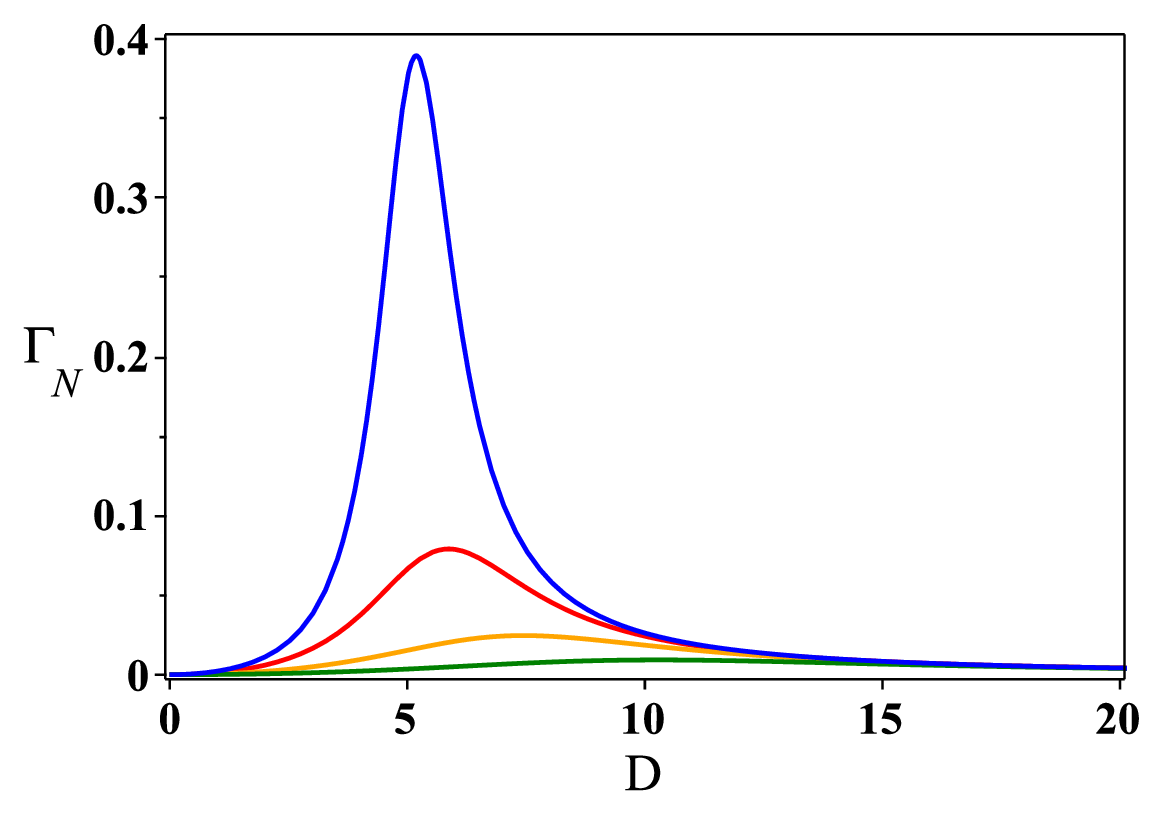}}
\end{center}
\caption{(Color online) The asymptotic transition rate,
$\Gamma_N$, as a function of the noise amplitude, $D$, for $\epsilon =5$,
$V=1$, $\gamma=1$, $\delta_a =1$ and $N=1,5,10,20$ (from up to
bottom).
\label{figGN}}
\end{figure}

\begin{figure}[tbh]
\begin{center}
\scalebox{0.65}{\includegraphics{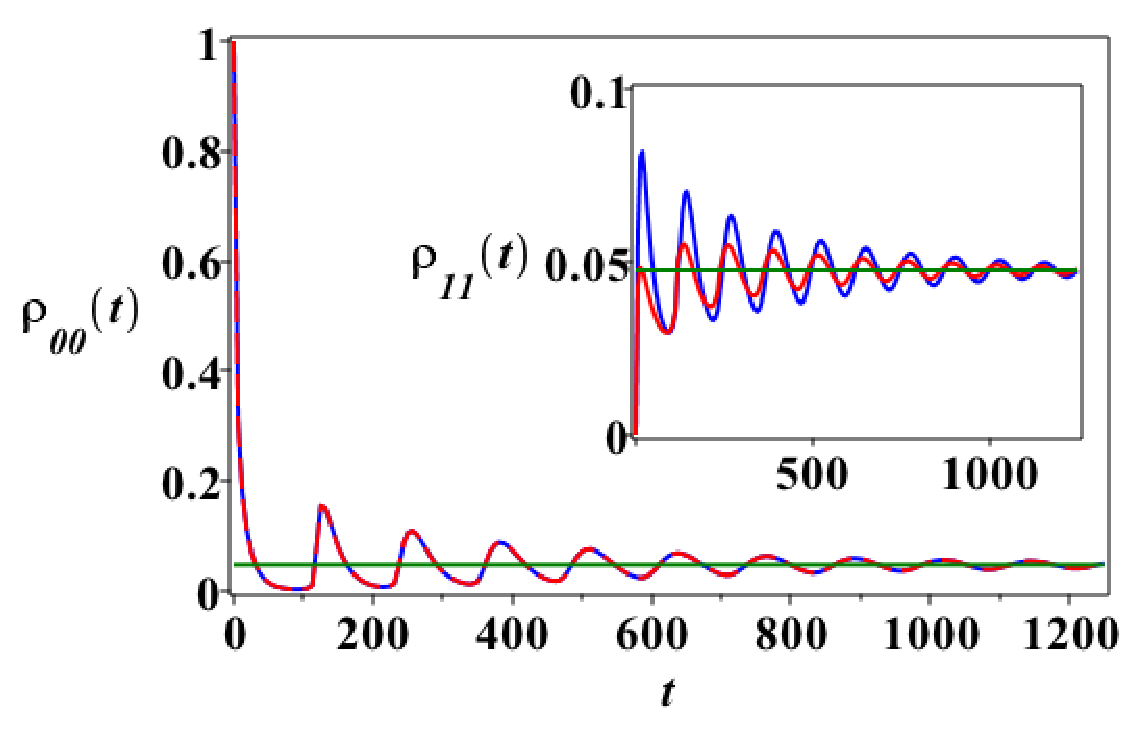}}
(a)
\scalebox{0.55}{\includegraphics{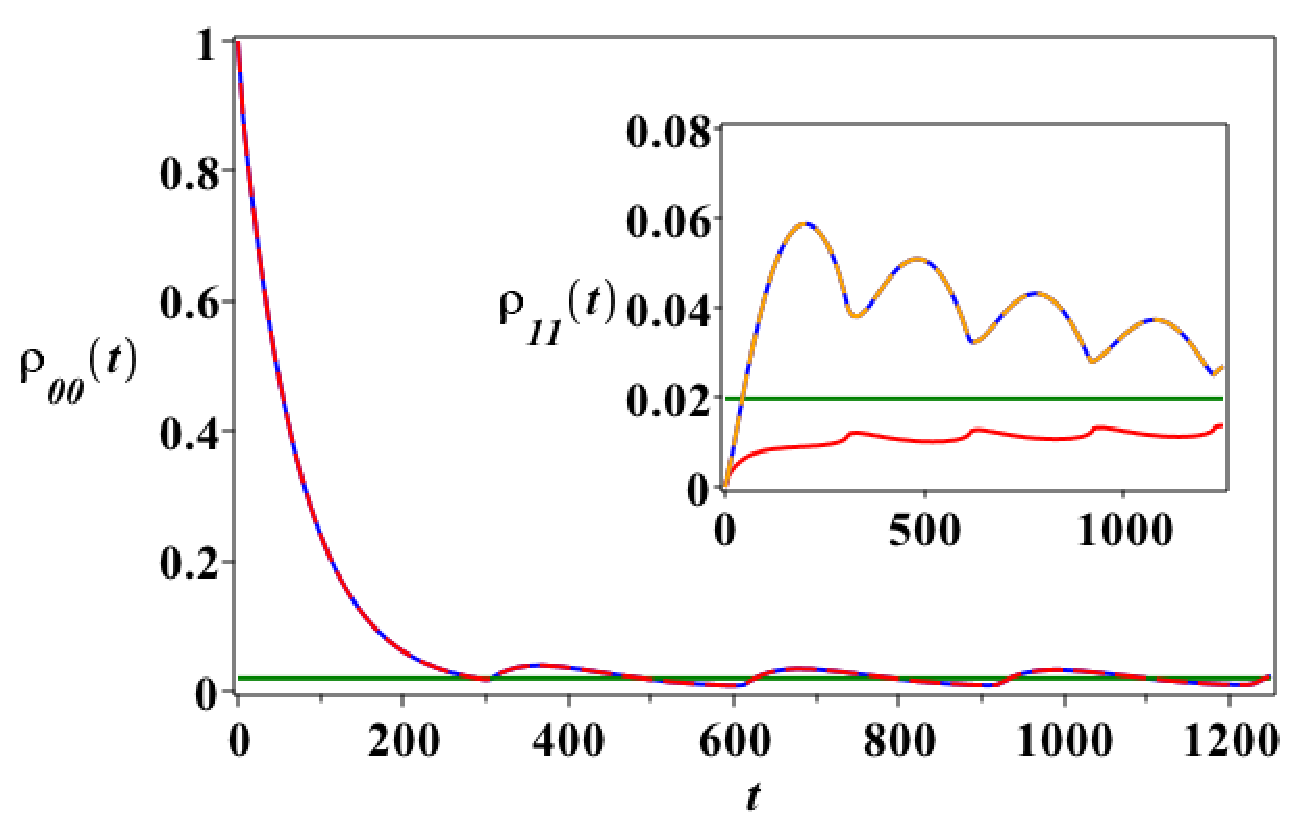}}
(b)
\end{center}
\caption{(Color online) Donor occupation as a function of time. The
asymptotic limit is shown by green line.
Parameters:  $V=1$, $D=5$, $\gamma=1$, $\delta_a =1$;
a) $N=20$, $E_0 =5$ (blue curve), $E_0 =-5$ (red curve).
Inset: occupation, $\rho_{11}(t)$, of level, $n=1$, of the acceptor.
(b) $N=50$, $E_0 =10$ (blue curve), $E_0 =-10$ (red curve).
Inset: occupation, $\rho_{11}(t)$, of level, $n=1$, of the acceptor. The
occupation, $\rho_{5050}(t)$, of level, $n=50$,  is depicted by orange
dashed curve.
\label{fig8}}
\end{figure}
The red and black curves correspond to $V=1$, $\gamma=1$ and $V=0.5$, $\gamma=0.25$, so that the ratio of $V^2/\gamma$ remains the same. These curves display the same transfer time at large $t$, in agreement with the scaling, Eq.~(\ref{scaling}). For comparison, we display $\rho_{00}^{}(t)$ for $V=0.5$, $\gamma =1$, which are out of the scaling (dashed-blue curve). This clearly shows a quite different transfer time.

In Fig. \ref{figGN}, the asymptotic transition rate,
$\Gamma_N=1/\tau_N$, as a function of the noise amplitude, $D$, is demonstrated for $N=1,5,10,20$. As it follows from Eq.(\ref{asocc4}), the transition rate, $\Gamma_N(D)$ experiences a resonant behavior with a maximum at $D=D_N$,
noise amplitude is in resonance with the redox potential, if
\begin{eqnarray}
D_N =\bigg((\epsilon^2 +4\gamma^2)\Big(\epsilon^2 +
\frac{V^4}{\delta_a^2 }
(N-1)^2\Big)\bigg)^{1/4}.
\end{eqnarray}
Using this results, one can rewrite Eq. (\ref{asocc4}) as,
\begin{eqnarray}
\tau_N^{}\propto \frac{D^4- 2\epsilon^2 D^2+ D_N^4}{\gamma  D^2 V^2}.
\label{asocc4g}
\end{eqnarray}
The transition time, $\tau_N$, reaches the minimum value at $D=D_N$, which corresponds to maximum value of the transition rate, $\Gamma_n$, in Fig. \ref{figGN},

\begin{eqnarray}
\tau_N^{min}=1/\Gamma_N^{max}\propto \frac{D_N^2- \epsilon^2  }{\gamma  V^2}.
\label{asocc4g}
\end{eqnarray}

\subsubsection*{Uphill ET:} In this case, the energy level of donor, $E_0$,
is positioned below the energy levels of the acceptor band. In
Fig.~\ref{fig8},
we presented the results of numerical simulations on comparison of downhill
and uphill ET. In Fig.~\ref{fig8}a, the blue curve demonstrates the downhill
ET ($E_0=5$), and the red curve demonstrates the uphill ET ($E_0=-5$), for
$\rho_{00}(t)$. In the insert the dynamics of $\rho_{11}(t)$ is shown for the
same parameters. The asymmetry in the behavior of blue and red curves
occurs because the energy level $E_1$ is the lowest level of the acceptor
band for the downhill ET (blue curve) and the upper level for the uphill ET
(red curve). In Fig.~\ref{fig8}b, we present similar results as in
Fig.~\ref{fig8}a, but for $N=50$ and $E_0=\pm 10$. As one can see from
the insert in Fig.~\ref{fig8}b, the symmetry is restored for the functions,
$\rho_{11}(t)$ (blue dashed curve) and $\rho_{50,50}(t)$ (orange dashed
curve).

Now, we will analyze the multi-scale time given by  Eq.~(\ref{asocc4}). It shows that the total depletion time of the donor $\tau_N^{}$ is proportional to $1/\delta_a^2$. Thus, $\tau_N^{}$ will strongly decrease with increasing bandwidth, $\delta_a$,   in particular when it approaches $\epsilon$. This is illustrated in Fig.~\ref{fig9}.
\begin{figure}[tbh]
\begin{center}
\scalebox{0.6}{\includegraphics{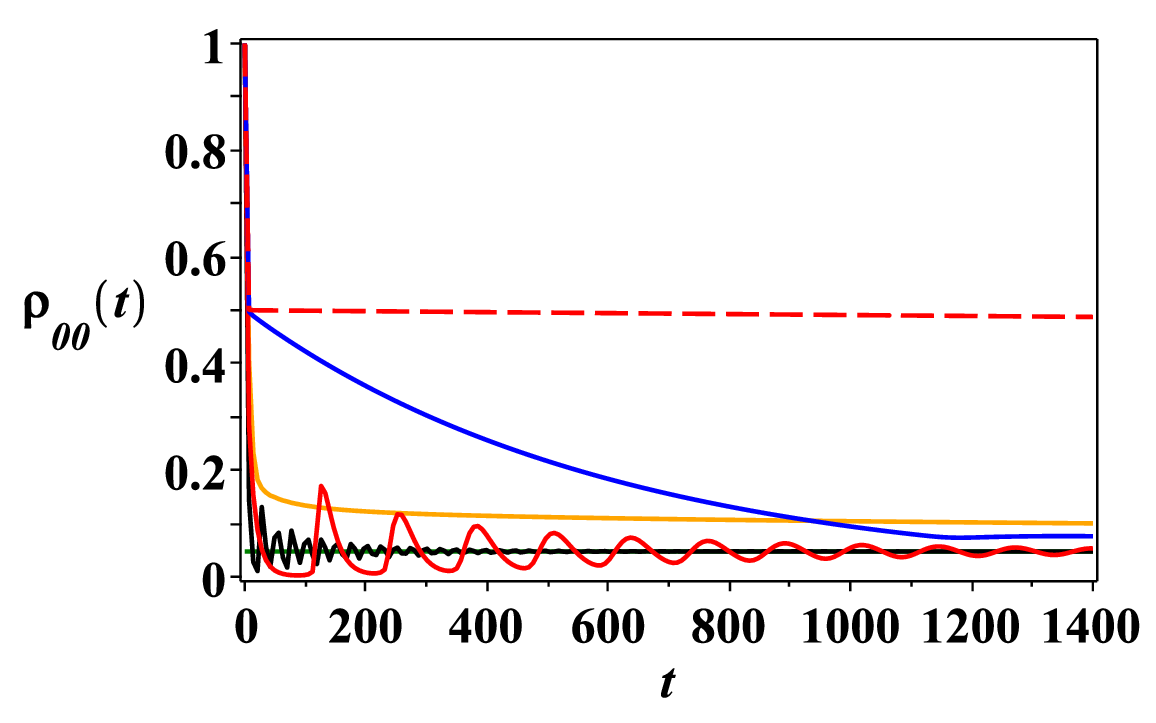}}
\end{center}
\caption {(Color online) Donor occupation as a function of time, for
$N=20$,
$\epsilon =D=5$, $V=\gamma =1$, and different values of the bandwidth:
$\delta_a = 40$ (orange), $ \delta_a = 5$ (black), $\delta_a = 1$ (red),
$\delta_a = 0.1$ (blue), $\delta_a = 0.01$ (red-dashed). The asymptotic
limit is shown by green line.
\label{fig9}}
\end{figure}
Here, too we used the exact equations Eqs.~(\ref{a15a})-(\ref{a15f}), for
evaluation of $\rho_{00}^{}(t)$.  As one can see from Fig.~\ref{fig9}, an
increase of the acceptor bandwidth, $\delta_a$, significantly decreases the
time of population of the acceptor.

\subsection{Non-equidistant acceptor' energy spectrum}

As one can see from the results shown in Figs.~\ref{fig5}, \ref{fig7}, and \ref{fig8}, the quantum coherent damping oscillations are observed during the process of the acceptor' population. In our model, these oscillations result from a re-population of the acceptor' states during the ET process, and they are especially pronounced under the condition: $\epsilon=D$. For the equidistant energy spectrum of the acceptor' band, given by  Eq.~(\ref{acen}), the period of these oscillations can be estimated as: $T=2\pi N/\delta$. The question arises if these quantum coherent oscillations will be observed for non-equidistant acceptor' energy spectrum, and up to what extent the destructive quantum interference effects could suppress them.

To clarify this issue, we performed numerical simulations for the non-equidistant acceptor' energy spectrum given by,
 \begin{eqnarray}
 E_n={2n-1-N\over 2(N-1)}\,\delta_a  +\frac{\kappa \delta_a}{N}\sin\bigg(\frac{\pi n}{\sqrt{3}}\bigg).
 \label{nonequid}
\end{eqnarray}
The first term in this expression describes the equidistant energy spectrum given by Eq.~(\ref{acen}), The second term in Eq.~(\ref{nonequid}) was chosen to model the non-equidistant part of the spectrum.
\begin{figure}[tbh]
\begin{center}
\scalebox{0.65}{\includegraphics{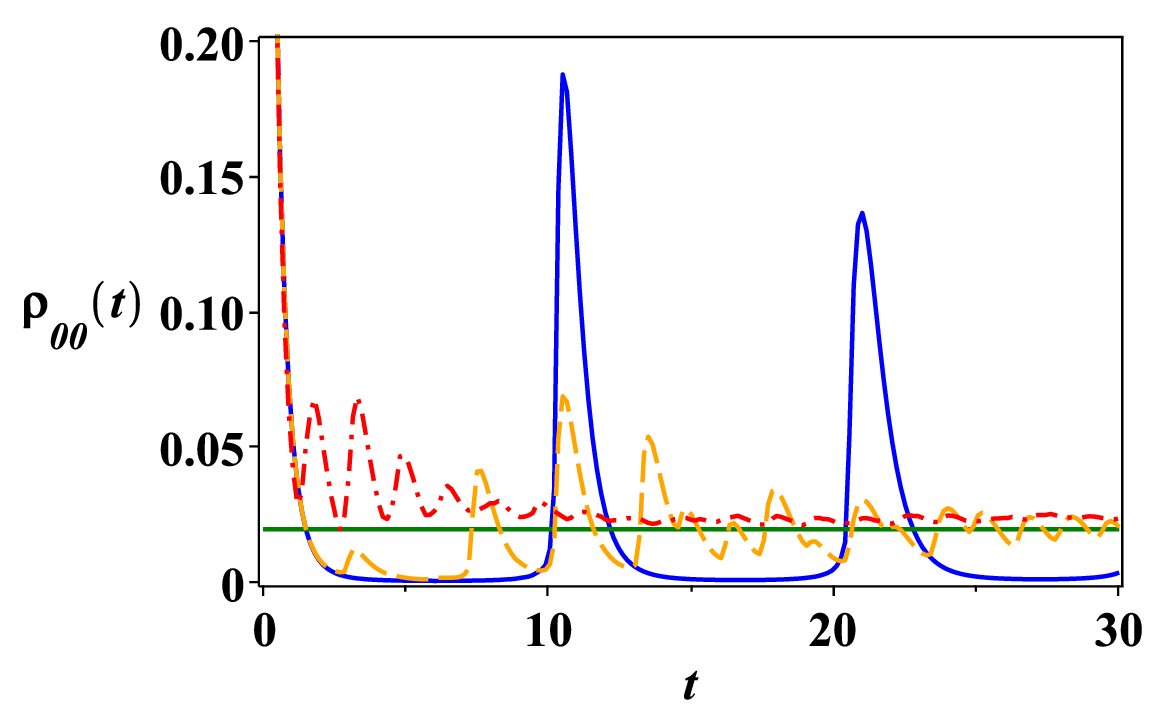}}
\end{center}
\caption{(Color online) Donor occupation as a function of time, for $N=50$,
$\delta_a =30$, $\epsilon =D=40$, $V= 10$,  $\gamma =10$, and
different values of the parameter $\kappa$: $\kappa = 0$ (blue), $\kappa =
0.25$ (orange dashed curve), $\kappa = 5$ (red dot-dashed curve). The
asymptotic limit is shown by green line.
\label{fig10B}}
\end{figure}
In Fig.~\ref{fig10B}, we present the results of numerical simulations of the donor population, for the non-equidistant energy spectrum of the acceptor band, Eq.~(\ref{nonequid}), and for different values of the parameter, $\kappa$, which describes the displacement from non-equidistance. The value, $\kappa=0$, corresponds to the equidistant energy spectrum (blue curve in  Fig.~\ref{fig10B}). One can see from Fig.~\ref{fig10B} that quantum coherent oscillations survive even for non-equidistant spectrum ($\kappa\neq 0$), if the non-equidistance is small enough. However, with increasing  $\kappa$}, these oscillations become quasi-periodic (involving many frequencies). The amplitudes of these oscillations become smaller, and the characteristic decay time decreases. These coherent quantum oscillations could be used for extracting characteristic spectroscopic  parameters in natural bio-complexes and for engineering artificial bio-nano devices.

\section{Conclusion}

We studied analytically and numerically noise-assisted quantum exciton (and electron) transfer (ET) in bio-complexes consisting of a single-level electron donor interacting with a multi-level acceptor. This situation takes place, in particular, when a
single excited energy level of the donor is populated, and it decays into a multi-level acceptor.
Our approach can also be used in a coarse-graining procedure, for a bio-complex with many energy levels considered as an electron ``donor" and/or``acceptor" with complicated internal structures. All  energy levels are assumed to interact  with  the protein-solvent environment, modeled by a diagonal classical noise, which corresponds (between other assumptions) to the high-temperature regime. Our approach can be applied for both the  exciton transfer in the light harvesting complexes and for the electron transfer in the reaction centers. We vary the number of the acceptor levels,  the acceptor bandwidth, the strength of the donor-acceptor interaction, and the noise amplitude and the correlation time. Under certain conditions, we derive  analytical expressions for the ET rate and efficiency.  We demonstrate that, for a relatively wide acceptor band, the efficiency of the ET from donor to acceptor can be close to 100\% for a broad range of noise amplitudes, for both sharp and flat redox potentials.
We show that generally the dynamics of the acceptor population can be characterized by multi-scale processes, which display a coarse-graining structure. We would like to note here that the multi-scale ET dynamics may result in decreasing the ET efficiency (probability of the acceptor population) due to such processes as fluorescence and  recombination of the  exciton through the interaction with the environment, which usually take place of the time-scales $\sim$ ns.

We also estimate the corresponding ET rates in a multi-scale regime. For our model, we obtain equal occupation of all levels at large times, independent of the structure of the acceptor band. This implies the possibility of optimizing  the efficiencies of the acceptor population by engineering the donor-acceptor complexes.

Our approach demonstrates a possibility of the efficient uphill population of the acceptor, due to the ``entropy factor" (large number of levels in the acceptor band). This result can be useful in many applications for controlling of the ET in bio-complexes. It would be important to experimentally verify this result, for example, in chlorophyll based heterodimer.

It is very remarkable that the coarse-graining dynamics for the electron transfer naturally occurs from our microscopic derivations, without any ad hoc assumptions.
Thus, our approach, developed in this paper, can also be applied as a consistent coarse-graining procedure that describes exciton and electron transfer in large bio-complexes. Indeed, the coarse-graining procedure usually suggests that connected clusters, which include many electron energy levels, are replaced by interacting effective ``donors" and ``acceptors", which are characterized by electron bandwidths with finite numbers of energy levels. Then, according to our results, the ET rates and efficiencies will strongly depend on the electron bandwidths of the corresponding clusters and on the number of electron energy levels localized in these clusters. By engineering in a controlled way the bandwidths of these clusters, one can significantly increase the efficiency of the ET in bio-complexes. To do this, one must take into account the structure of energy levels and their interactions inside the individual bio- clusters; interactions between different clusters; and their interactions with local and collective environments. The quantum coherent oscillations, which we discussed, are analogous to the quantum coherent effects observed in experiments on exciton transfer in photosynthetic complexes, such as the Fenna-Matthews-Olson (FMO) bacteriochlorophyll complex, which is found in green sulphur bacteria \cite{ECR}, and in photosynthetic marine algae \cite{CWWC}, at ambient temperature. These oscillations could be used in the spectroscopic experiments, such as dynamic fluorescence, for resolving the structures of energy spectra inside the bio-complexes. At the same time, we should mention that presently, there is no consensus about the origin of these oscillations, as they can have even completely classical vibrational origin. All this will require further analytical, numerical, and experimental studies.

\section*{Acknowledgments}

    We are thankful to A. Aharony, O. Entin-Wohlman, and G.D. Doolen for
    useful comments. S.G. acknowledges the Beijing Computational Science Research Center for supporting his visit, where a part of this work was done. A.I.N. acknowledges the support from the CONACyT. G.P.B. thanks the Einstein Center and the Goldschleger Center for Nanophysics, at the Weizmann Institute, for supporting his visit to Israel, where a part of this work was done.

\appendix

\section{}

We consider the system of algebraic equations (\ref{d2}), written in the form:
\begin{eqnarray}
\sum_{n=1}^N{\cal A}_{mn}^{(N)}X_{n}=0,
\end{eqnarray}
where ${\cal A}_{mn}^{(N)}= \delta_{mn} - A_{mn}$ and,
\begin{eqnarray}
 A_{mn} = \frac{C_{mn}}{1+ F_m}.
 \label{A1}
\end{eqnarray}
Here, $C_{mn}=V_N^2/(\epsilon_{mn}^2+4\gamma^2)$  and $F_m=\sum_{n}C_{mn}$.

Let  $\lambda_i$ be the eigenvalues of the matrix, $A$. Employing the Perron-Frobenius theorem \cite{X}, one can show that,
\begin{equation}
|\lambda_i| \leq {\max}\big\{\sum_n A_{mn}\big\} = \max \bigg\{\frac{F_m}{1+F_m}\bigg\} <1.
\end{equation}

 To evaluate ${\rm det}({\mathbb I}-B)$, we use the following formula:
 \begin{eqnarray}
  {\rm det}({\mathbb I}-A) = \exp\big( {\rm Tr}\ln({\mathbb I}-A)\big).
 \end{eqnarray}
Expanding the logarithm in the series, we find,
\begin{eqnarray}
 {\rm det}\,{\cal A}^{(N)} =  {\rm det}({\mathbb I}-A) = \exp\bigg( -\sum^\infty_{n=1} \frac{1}{n} {\rm tr} A^n \bigg).
 \label{A2}
 \end{eqnarray}
Then, using the equation, ${\rm tr} A^n = \sum^N_{i=1}\lambda^n_i$, we obtain,
\begin{eqnarray}
 {\rm det}\,{\cal A}^{(N)} =\exp\bigg( -\sum^N_i\sum^\infty_{n=1} \frac{\lambda^n_i}{n} \bigg).
 \label{A2}
 \end{eqnarray}
From here it follows,
\begin{eqnarray}
 {\rm det}\,{\cal A}^{(N)} \geq\exp\bigg( -\sum^N_i\sum^\infty_{n=1} \frac{|\lambda_i|^n}{n} \bigg). \label{A2}
 \end{eqnarray}
One can recast inequality (\ref{A2}) as,
\begin{eqnarray}
 {\rm det}\,{\cal A}^{(N)} \geq\prod^N_{i=1}\big( 1-|\lambda_i|\big) >0.
 \label{A3}
 \end{eqnarray}

Let $F_0$ be the value of $F_m$ ($m=1,2,\dots N$) yielding the maximum of $\sum_n A_{mn}$,
\begin{eqnarray}
{\max}\big\{\sum_n A_{mn}\big\} = \frac{F_0}{1+F_0}.
\end{eqnarray}
Then, the estimate (\ref{A3}) can be simplified as follows:
\begin{eqnarray}
 {\rm det}\,{\cal A}^{(N)} \geq\frac{1}{(1+F_0)^N} .
 \label{A4}
 \end{eqnarray}

\end{document}